\documentclass[fleqn,usenatbib]{mnras}

\usepackage{newtxtext,newtxmath}

\usepackage[T1]{fontenc}
\usepackage{ae,aecompl}

\usepackage{graphicx}	
\usepackage{amsmath}	
\usepackage{amssymb}	

\title[The SEDs of AGNs]{The Spectral Energy Distributions of Active Galactic Nuclei}

\author[M. J. I. Brown et al.]{M. J. I .Brown,$^{1}$\thanks{E-mail: Michael.Brown@Monash.edu}
K. J. Duncan,$^{2}$
H. Landt,$^{3}$
M. Kirk$,^{1}$
C. Ricci,$^{4,5,6}$
N. Kamraj,$^{7}$  \newauthor
M. Salvato,$^{8}$
T. Ananna$^{9,10}$
\\
$^{1}$School of Physics \& Astronomy, Monash University, Clayton, Victoria 3800, Australia\\
$^{2}$Leiden Observatory, Leiden University, NL-2300 RA Leiden, Netherlands\\
$^{3}$Department of Physics, Centre for Extragalactic Astronomy, Durham University, South Road, Durham DH1 3LE, UK\\
$^{4}$N\'ucleo de Astronom\'ia de la Facultad de Ingenier\'ia, Universidad Diego Portales, Av. Ej\'ercito Libertador 441, Santiago, Chile\\
$^{5}$Kavli Institute for Astronomy and Astrophysics, Peking University, Beijing 100871, China\\
$^{6}$Chinese Academy of Sciences South America Center for Astronomy, Camino El Observatorio 1515, Las Condes, Santiago, Chile\\
$^{7}$Cahill Center for Astronomy and Astrophysics, California Institute of Technology, Pasadena, CA 91125, USA\\ 
$^{8}$MPE, Giessenbachstrasse 1, Garching 85748, Germany\\
$^{9}$Department of Physics, Yale University, P.O. Box 201820, New Haven, CT 06520-8120, USA\\
$^{10}$Yale Center for Astronomy and Astrophysics, P.O. Box 208121, New Haven, CT 06520, USA\\ }

\date{Accepted XXX. Received YYY; in original form ZZZ}

\pubyear{2018}

\hypersetup{draft} 

\begin{document}
\label{firstpage}
\pagerange{\pageref{firstpage}--\pageref{lastpage}}
\maketitle

\begin{abstract}
We present spectral energy distributions (SEDs) of 41 active galactic nuclei, derived from multiwavelength photometry and archival spectroscopy. All of the SEDs span at least 0.09 to $30~{\rm \mu m}$, but in some instances wavelength coverage extends into the X-ray, far-infrared and radio. For some AGNs we have fitted the measured far-infrared photometry with greybody models, while radio flux density measurements have been approximated by power-laws or polynomials. We have been able to fill some of the gaps in the spectral coverage using interpolation or extrapolation of simple models. In addition to the 41 individual AGN SEDs, we have produced 72 Seyfert SEDs by mixing SEDs of the central regions of Seyferts with galaxy SEDs.  Relative to the literature, our templates have broader wavelength coverage and/or higher spectral resolution. We have tested the utility of our SEDs by using them to generate photometric redshifts for $0<z \leq 6.12$ AGNs in the Bo\"otes field (selected with X-ray, IR and optical criteria) and, relative to SEDs from the literature, they produce comparable or better photometric redshifts with reduced flux density residuals. 
\end{abstract}

\begin{keywords}
galaxies: active -- (galaxies:) quasars: general -- galaxies: Seyfert -- (galaxies:) quasars: emission lines -- galaxies: distances and redshifts
\end{keywords}



\section{Introduction}

Templates or models of active galactic nucleus (AGN) spectral energy distributions (SEDs) can be critical for k-corrections,  exposure time calculations, modelling quasar selection,  modelling the emission from quasars and photometric redshifts, \citep[e.g. see the review of AGN photometric redshifts of ][]{salvato2018}. While AGN SEDs are sometimes described as power-laws, there are several continuum features that can impact broadband photometry, including the big blue bump, a $\sim 1~{\rm \mu m}$ inflection attributed to changing contributions from the disk and torus, and mid-infrared features from silicate emission \citep[e.g.,][ and references therein]{elvis1994,haa05,shang2005,netzer2007}. Emission lines can also be sufficiently strong in Type 1 AGNs to impact broadband photometry. For example, at $z>3.5$ the broad ${\rm H\alpha}$ emission of Type 1 quasars can shift their observed colours towards and away from the locus of galaxy colours \citep[e.g.,][]{ric01,cool2006}. While complicating quasar selection and k-corrections, these spectral features are of utility for photometric redshifts. 

\begin{figure*}
 \includegraphics[width=0.8\textwidth]{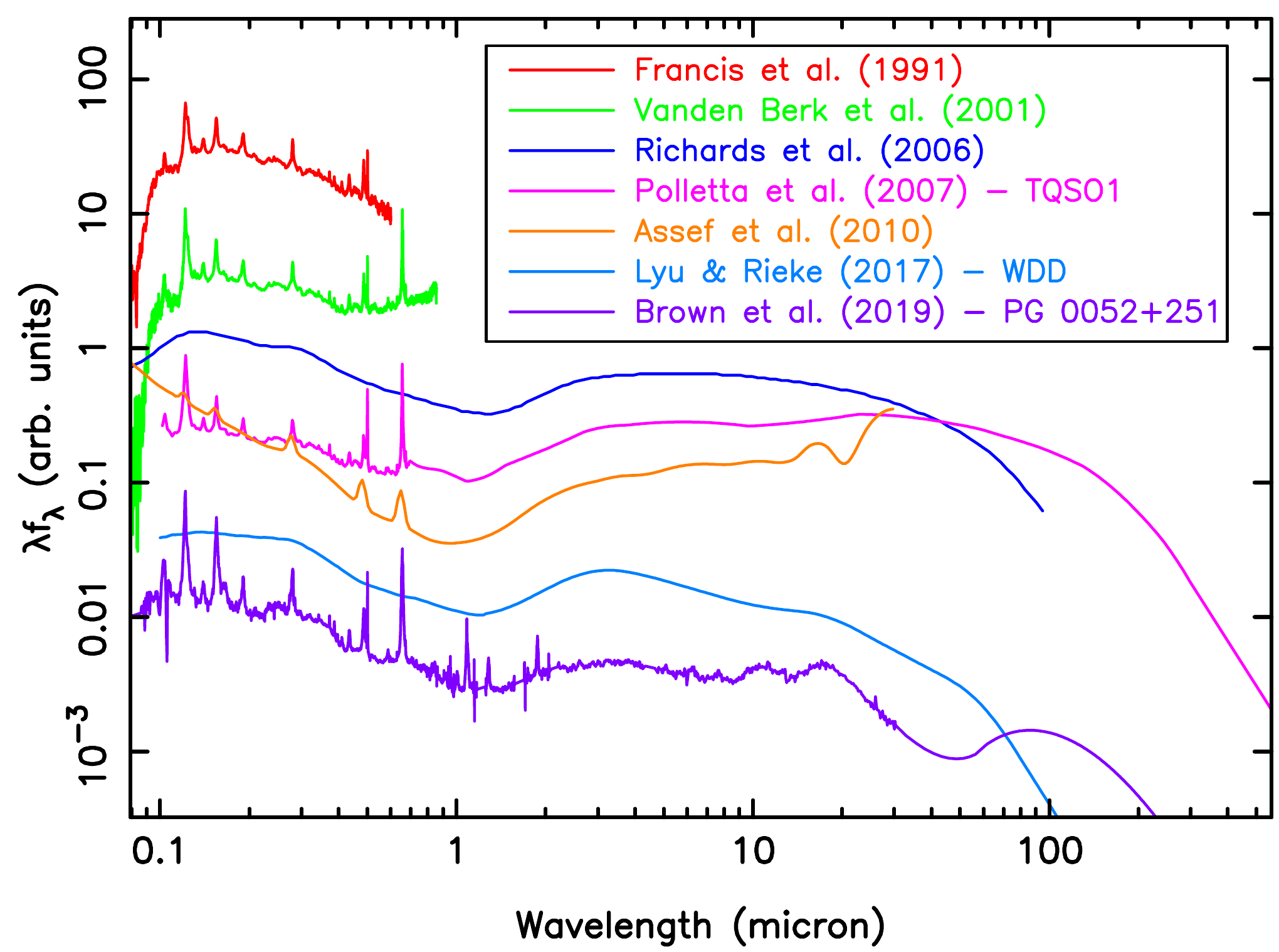}
 \caption{Illustrative examples of quasar spectral energy distributions from the past two decades \citep{fra91,vandenberk01,ric06,pol07,assef2010,lyu2017b}, along with our SED for PG 0052+251. These templates were created using different methods and with different purposes, so some caution is required directly comparing them. However, the trend towards improved wavelength coverage and spectral resolution is evident. Apart from the improved spectral resolution, our SED for PG 0052+251 shows prominent mid-infrared silicate emission features \citep[also see][]{netzer2007} and a peak at $\sim 100~{\rm \mu m}$ caused by thermal emission by dust.}
 \label{fig:iauqsosed}
\end{figure*}

AGN SEDs with broad wavelength coverage have been created by combining spectroscopy and/or photometry of many individual quasars, and exploiting the broad observed redshift range of quasars \citep[e.g.,][]{fra91,elvis1994,vandenberk01,pol07,salvato2009,assef2010,hsu2014,sel16}. As Figure~\ref{fig:iauqsosed} illustrates, this approach can produce high signal-to-noise mean spectra and principal components. A risk is the composite spectra may not correspond to any individual quasar, and the principal components can produce unphysical spectra (i.e., when used to model noisy photometry), although in practice such SEDs can approximate individual quasars. SEDs produced using photometry of many quasars understandably have low resolution, and this can result in significant spectral features being absent from AGN SED templates. For example, the diversity seen in {\it Spitzer} IRS mid-infrared spectroscopy of quasars \citep[e.g.,][]{haa05,netzer2007,lyu2017a} is largely absent from composite quasar SEDs, with most SEDs in Figure~\ref{fig:iauqsosed} not showing the two silicate emission features seen in the SED of PG~0052+251.

The alternative to composite AGN SED templates is photometry and spectrophotometry of individual objects. This approach has been used for over 50 years, with the wavelength range, precision and spectral resolution of AGN SEDs improving over that time \citep[e.g.,][]{oke63,oke70,deb78,edelson1986,shang2005,her16}. Spectrophotometry of individual AGNs has expanded from optical spectrophotometry into the UV, near-IR and mid-IR \citep[e.g.,][]{shang2005,landt2008,her16}. However, while spectrophotometry does exist of individual quasars between $0.1$ and $30~{\rm \mu m}$, there have been no SEDs of individual quasars that use spectrophotometry to span this whole wavelength range \citep[although photometric SEDs with broad wavelength coverage have been available for decades; e.g.,][]{edelson1986,elvis1994,spinoglio2002}. A significant hurdle is the lack of spectrophotometry of AGNs in the observed $1$ to $2.5~{\rm \mu m}$ range, although there are some notable exceptions \citep[e.g.,][]{glikman2006,riffel2006,landt2008,landt2013}.

Relative to spectrophotometry of galaxies, the construction of AGN SEDs offers some advantages and challenges. For sufficiently luminous quasars, emission across much of the electromagnetic spectrum is effectively from a point source, mitigating issues associated with varying (extraction) aperture bias as a function of wavelength. However, quasar variability means spectra taken years (or months) apart may not measure the same flux density at a given wavelength, and the strength and width of emission lines may have changed \citep[e.g.,][]{matthews1963,hook1994,soldi2008,simm2016}. While stellar population models \citep[e.g.,][]{bc03,dac08} can be used to cover gaps in spectral coverage for galaxies \citep[e.g.,][]{bro14}, quasar spectra have emission lines and continuum features that aren't always trivial to model (particularly if photometry as a function of wavelength is impacted by variability).

In this paper we present SEDs of 41 individual AGNs created by combining archival spectrophotometry spanning broad wavelength ranges\footnote{The observed and restframe SEDs are available via DOI 10.17909/t9-3dbt-8734.}. All of the AGN SEDs span $0.09$ to $30~{\rm \mu m}$, but in some instances SEDs have been expanded into the X-ray, far-infrared and radio. Our goal is to provide SEDs that have higher resolution and greater precision than existing libraries. However, clear risks include variability and wavelength dependent aperture bias introducing wavelength dependent errors. We thus test the utility of our SEDs to produce accurate photometric redshifts and quantify their performance using several metrics (e.g. redshift errors and flux density residuals).

The structure of this paper is as follows. In Section~\ref{sec:datamodels} we describe the input spectroscopy, photometry and models used for the SEDs. We describe the construction of the SEDs in Section~\ref{sec:construction}, including rescaling the input spectra and models to produce continuous SEDs, and the production of Seyfert SEDs using combinations of AGN and host galaxy SEDs. In Section~\ref{sec:colours} we provide an overview of the SEDs, including their restframe colours and observed colours as a function of redshift. We discuss the utility of the SEDs for photometric redshifts in Section~\ref{sec:photoz}, including comparison with other SED template libraries and quantification of photometric redshift performance (e.g. flux density residuals). We provide the principal conclusions of this work in Section~\ref{sec:conclusions}. Throughout this paper we use AB magnitudes and a flat cosmology with $\Omega_m=0.3$ and $H_0=70~{\rm km~s^{-1}~Mpc^{-1}}$. 

\section{Input data and models}
\label{sec:datamodels}

\subsection{Spectrophotometry}

Our AGN sample was selected based on the availability of spectrophotometry spanning the 0.1 and $30~{\rm \mu m}$ wavelength range, although some AGNs have gaps in this wavelength range while others have X-ray and far-infrared spectra. We draw heavily on ground-based optical and near-infrared spectroscopy from \citet{shang2005}, \citet{riffel2006}, \citet{landt2008} and \citet{landt2013}. At other wavelengths we utilise spectrophotometry from the Nuclear Spectroscopic Telescope Array (NuSTAR), {\it Suzaku}, {\it XMM-Newton}, the {\it Chandra} X-ray observatory, the Far Ultraviolet Spectroscopic Explorer (FUSE), the Hopkins Ultraviolet Telescope (HUT), the International Ultraviolet Explorer (IUE), the Wisconsin Ultraviolet Photo-Polarimeter Experiment (WUPPE), the Galaxy Evolution Explorer (GALEX), {\it Hubble} Space Telescope (HST), {\it Akari}, {\it Spitzer}, the Infrared Space Observatory (ISO) and {\it Herschel}. Given our extensive use of archival spectra, our sample has significant overlap with existing studies and compilations of AGN spectra and SEDs including AGN Watch \citep{peterson1999}, FOSAGN \citep{evans2004} and studies exploiting {\it Akari} and {\it Spitzer} spectroscopy \cite[e.g.,][]{shi2014,kim2015}. 

Selecting AGNs on the basis of available multi-wavelength spectrophotometry results in a sample biased towards relatively bright and well studied broadline AGNs. Furthermore, as we want to create continuous spectrophotometry taken with a variety of instruments, we tend to favour bright quasars or nearby galaxy nuclei that are dominated by AGN light, as aperture bias hinders making SEDs of objects where the spectra are complex blends of AGN and host galaxy light. The issue of aperture  bias is highlighted by the variety of extraction apertures used, which include the IUE $20^{\prime\prime} \times 10^{\prime\prime}$ elliptical aperture, HST STIS slits that vary between 0.05 to $2^{\prime\prime}$ width, the SDSS $3^{\prime\prime}$ diameter fibre, and {\it Spitzer} IRS low resolution spectroscopy slits that are $3.6^{\prime\prime}$ and $10.6^{\prime\prime}$ wide (short and long of $14~{\rm \mu m}$). 

Our requirement for multi-wavelength spectrophotometry biases us against obscured AGNs that mostly lack ultraviolet spectrophotometry and/or have significant host galaxy contributions to the spectrum at wavelengths short of $3~{\rm \mu m}$. To mitigate but not eliminate this bias we include some obscured AGNs that lack spectrophotometry between 0.1 and $30~{\rm \mu m}$, replacing the missing spectrophotometry with simple empirical models. 

Our final sample consists of 41 AGNs, whose basic properties are summarised in Table~\ref{table:agnsummary}. Our use of archival data results in us using spectra from a large number of archives and papers that are detailed in Table~\ref{table:specsummary}. In some instances we have used the spectra as published in papers whereas in other instances we have used more recent reductions provided by archives including the Combined Atlas of Sources with {\it Spitzer} IRS Spectra \citep[CASSIS;][]{leb11}, the Barbara A. Mikulski Archive for Space Telescopes (MAST) and the Hubble Spectral Legacy Archive \citep[HSLA; ][]{peeples2017}

\begin{table*}
	\centering
	\caption{The basic properties of the sample and the SED wavelength ranges.}
	\label{table:agnsummary}
    \scriptsize
	\begin{tabular}{lrrcllllllll} 
		\hline
Name & \multicolumn{2}{c}{Coordinates} & $z$ & $m_g$ & $M_g$ & $E(B-V)$ & Phot. & AGN & ${\rm EW(H\alpha+[NII])}$ & SED $\lambda$ & Scale \\
    &  \multicolumn{2}{c}{(J2000)}     &     &       &       & (mag)    & Aper. & Class$^\dagger$ & (\AA) & range (${\rm \mu m}$) & range \\
		\hline
2MASX J130005...   & 195.0223 &  16.5374 & 0.0799 & 17.0 & -20.8 &  0.02 & $ 10^{\prime\prime}$  &  S1i   &   62  & $ 9.0 \times 10^{-2} - 6.4 \times 10^{1} $  & $ 1.0 - 2.5 $ \\ 
             3C 120   &  68.2962 &   5.3543 & 0.0330 & 13.5 & -21.8 &  0.36 & $ 10^{\prime\prime}$  &  S1.5  &  817  & $ 1.0 \times 10^{-5} - 4.8 \times 10^{6} $  & $ 1.0 - 1.7 $ \\ 
             3C 273   & 187.2779 &   2.0524 & 0.1583 & 12.8 & -26.4 &  0.03 & $ 10^{\prime\prime}$  &  S1.0  &  446  & $ 9.0 \times 10^{-6} - 3.4 \times 10^{6} $  & $ 0.9 - 1.2 $ \\ 
             3C 351   & 256.1724 &  60.7418 & 0.3719 & 15.3 & -25.8 &  0.02 & $ 10^{\prime\prime}$  &  S1.5  &  349  & $ 9.9 \times 10^{-5} - 7.1 \times 10^{6} $  & $ 0.5 - 1.4 $ \\ 
           3C 390.3   & 280.5375 &  79.7714 & 0.0561 & 14.9 & -21.7 &  0.09 & $ 10^{\prime\prime}$  &  S1.5  & 1441  & $ 1.0 \times 10^{-5} - 9.5 \times 10^{6} $  & $ 0.5 - 25 $ \\ 
            Ark 120   &  79.0475 &  -0.1498 & 0.0327 & 13.6 & -22.1 &  0.12 & $ 10^{\prime\prime}$  &  S1.0  &  529  & $ 2.0 \times 10^{-5} - 2.9 \times 10^{5} $  & $ 0.8 - 4.0 $ \\ 
            Ark 564   & 340.6639 &  29.7254 & 0.0247 & 14.5 & -20.5 &  0.06 & $ 10^{\prime\prime}$  &  S3    &  197  & $ 8.8 \times 10^{-2} - 3.6 \times 10^{1} $  & $ 0.8 - 1.2 $ \\ 
          Fairall 9   &  20.9407 & -58.8058 & 0.0470 & 13.9 & -22.3 &  0.02 & $ 10^{\prime\prime}$  &  S1.2  &  574  & $ 2.5 \times 10^{-5} - 9.6 \times 10^{2} $  & $ 0.6 - 2.1 $ \\ 
       F2M1113+1244   & 168.4777 &  12.7443 & 0.6812 & 20.7 & -23.4 &  0.02 & $ 10^{\prime\prime}$  &  S1    &  430  & $ 9.0 \times 10^{-2} - 3.5 \times 10^{1} $  & $ 0.5 - 1.0 $ \\ 
         H 1821+643   & 275.4888 &  64.3434 & 0.2968 & 14.2 & -26.6 &  0.05 & $ 10^{\prime\prime}$  &  S1.2  &  636  & $ 7.0 \times 10^{-2} - 1.5 \times 10^{6} $  & $ 1.0 - 1.2 $ \\ 
    IRAS 11119+3257   & 168.6620 &  32.6926 & 0.1876 & 18.6 & -21.8 &  0.02 & $ 10^{\prime\prime}$  &  S1n   &  370  & $ 9.0 \times 10^{-2} - 4.2 \times 10^{2} $  & $ 1.0 - 1.2 $ \\ 
   IRAS F16156+0146   & 244.5392 &   1.6559 & 0.1320 & 18.5 & -20.3 &  0.08 & $ 10^{\prime\prime}$  &  S2    & 511   & $ 9.0 \times 10^{-2} - 3.3 \times 10^{1} $  & $ 0.9 - 2.3 $ \\ 
            Mrk 110   & 141.3035 &  52.2862 & 0.0353 & 15.8 & -21.0 &  0.01 & $ 10^{\prime\prime}$  &  S1n   &  917  & $ 2.0 \times 10^{-5} - 3.6 \times 10^{1} $  & $ 0.6 - 8.0 $ \\ 
           Mrk 1502   &  13.3956 &  12.6934 & 0.0589 & 14.2 & -22.8 &  0.08 & $ 10^{\prime\prime}$  &  S1n   &  346  & $ 8.5 \times 10^{-2} - 2.9 \times 10^{5} $  & $ 0.6 - 2.0 $ \\ 
            Mrk 231   & 194.0593 &  56.8737 & 0.0422 & 13.7 & -22.4 &  0.01 & $ 10^{\prime\prime}$  &  S1.0  &  293  & $ 4.4 \times 10^{-5} - 9.6 \times 10^{6} $  & $ 0.9 - 2.0 $ \\ 
            Mrk 279   & 208.2644 &  69.3082 & 0.0305 & 14.3 & -21.6 &  0.02 & $ 10^{\prime\prime}$  &  S1.0  &  242  & $ 1.0 \times 10^{-5} - 9.7 \times 10^{2} $  & $ 0.9 - 2.1 $ \\ 
            Mrk 290   & 233.9683 &  57.9026 & 0.0302 & 15.0 & -20.1 &  0.01 & $ 10^{\prime\prime}$  &  S1.5  &  303  & $ 1.0 \times 10^{-5} - 3.6 \times 10^{1} $  & $ 1.0 - 1.1 $ \\ 
            Mrk 421   & 166.1138 &  38.2088 & 0.0300 & 13.4 & -22.3 &  0.02 & $ 10^{\prime\prime}$  &  HP    & -     & $ 1.0 \times 10^{-5} - 3.6 \times 10^{1} $  & $ 1.0 - 2.9 $ \\ 
            Mrk 493   & 239.7901 &  35.0299 & 0.0310 & 15.2 & -20.3 &  0.03 & $ 10^{\prime\prime}$  &  S1n   &  226  & $ 8.7 \times 10^{-2} - 3.6 \times 10^{1} $  & $ 1.0 - 2.0 $ \\ 
            Mrk 509   & 311.0406 & -10.7235 & 0.0344 & 13.2 & -22.3 &  0.06 & $ 15^{\prime\prime}$  &  S1.5  &  621  & $ 1.0 \times 10^{-5} - 9.7 \times 10^{2} $  & $ 1.0 - 1.2 $ \\ 
            Mrk 590   &  33.6398 &  -0.7667 & 0.0261 & 14.8 & -19.9 &  0.03 & $ 10^{\prime\prime}$  &  S1.0  &   43  & $ 4.2 \times 10^{-5} - 9.8 \times 10^{2} $  & $ 0.2 - 1.5 $ \\ 
            Mrk 817   & 219.0919 &  58.7943 & 0.0315 & 14.3 & -21.4 &  0.01 & $ 10^{\prime\prime}$  &  S1.5  &  406  & $ 2.6 \times 10^{-5} - 1.4 \times 10^{6} $  & $ 0.6 - 1.2 $ \\ 
            Mrk 876   & 243.4882 &  65.7193 & 0.1290 & 14.6 & -23.9 &  0.03 & $ 20^{\prime\prime}$  &  S1.0  &  789  & $ 9.0 \times 10^{-6} - 1.3 \times 10^{6} $  & $ 0.6 - 1.5 $ \\ 
            Mrk 926   & 346.1812 &  -8.6857 & 0.0469 & 15.0 & -21.8 &  0.04 & $ 10^{\prime\prime}$  &  S1.5  &  521  & $ 2.0 \times 10^{-5} - 9.6 \times 10^{2} $  & $ 0.5 - 3.0 $ \\ 
   NGC 3227 Central   & 155.8774 &  19.8651 & 0.0039 & 13.4 & -17.1 &  0.04 & $ 10^{\prime\prime}$  &  S1.5  &  547  & $ 2.8 \times 10^{-5} - 1.0 \times 10^{3} $  & $ 0.3 - 2.0 $ \\ 
   NGC 3516 Central   & 166.6979 &  72.5686 & 0.0088 & 13.3 & -19.0 &  0.04 & $ 10^{\prime\prime}$  &  S1.5  &  218  & $ 1.0 \times 10^{-5} - 9.9 \times 10^{2} $  & $ 0.8 - 2.0 $ \\ 
   NGC 4051 Central   & 180.7901 &  44.5313 & 0.0023 & 13.6 & -15.8 &  0.02 & $ 10^{\prime\prime}$  &  S1n   &  182  & $ 1.0 \times 10^{-5} - 6.9 \times 10^{1} $  & $ 0.7 - 1.2 $ \\ 
   NGC 4151 Central   & 182.6357 &  39.4057 & 0.0033 & 12.9 & -18.5 &  0.02 & $ 10^{\prime\prime}$  &  S1.5  &  640  & $ 2.1 \times 10^{-5} - 1.0 \times 10^{3} $  & $ 0.8 - 1.7 $ \\ 
   NGC 5548 Central   & 214.4981 &  25.1368 & 0.0166 & 14.5 & -19.3 &  0.02 & $ 10^{\prime\prime}$  &  S1.5  &  241  & $ 2.7 \times 10^{-5} - 9.9 \times 10^{2} $  & $ 0.2 - 1.1 $ \\ 
           NGC 5728   & 220.5996 & -17.2531 & 0.0094 & 13.3 & -19.7 &  0.12 & $ 15^{\prime\prime}$  &  S1.9  &   60  & $ 1.0 \times 10^{-5} - 9.9 \times 10^{2} $  & $ 0.2 - 4.5 $ \\ 
           NGC 7469   & 345.8151 &   8.8739 & 0.0163 & 13.3 & -20.6 &  0.08 & $ 15^{\prime\prime}$  &  S1.5  &  289  & $ 3.0 \times 10^{-5} - 9.9 \times 10^{2} $  & $ 0.6 - 1.9 $ \\ 
             OQ 530   & 214.9441 &  54.3874 & 0.1525 & 15.7 & -23.5 &  0.03 & $ 10^{\prime\prime}$  &  HP    & -     & $ 7.8 \times 10^{-2} - 6.9 \times 10^{1} $  & $ 0.7 - 1.0 $ \\ 
        PG 0026+129   &   7.3071 &  13.2677 & 0.1420 & 14.9 & -24.1 &  0.08 & $ 10^{\prime\prime}$  &  S1.2  &  320  & $ 1.3 \times 10^{-5} - 3.3 \times 10^{1} $  & $ 1.0 - 2.5 $ \\ 
        PG 0052+251   &  13.7172 &  25.4275 & 0.1545 & 15.4 & -23.8 &  0.06 & $ 10^{\prime\prime}$  &  S1.2  &  603  & $ 9.0 \times 10^{-6} - 8.7 \times 10^{2} $  & $ 0.9 - 1.6 $ \\ 
        PG 1211+143   & 183.5736 &  14.0536 & 0.0809 & 14.2 & -23.4 &  0.03 & $ 10^{\prime\prime}$  &  S1n   &  588  & $ 8.5 \times 10^{-2} - 9.3 \times 10^{2} $  & $ 0.5 - 1.6 $ \\ 
        PG 1307+085   & 197.4454 &   8.3302 & 0.1538 & 15.4 & -23.5 &  0.04 & $ 10^{\prime\prime}$  &  S1.2  &  597  & $ 7.9 \times 10^{-2} - 8.7 \times 10^{2} $  & $ 0.5 - 1.0 $ \\ 
        PG 1415+451   & 214.2534 &  44.9351 & 0.1137 & 16.2 & -22.4 &  0.01 & $ 10^{\prime\prime}$  &  S1.0  &  297  & $ 8.3 \times 10^{-2} - 3.4 \times 10^{1} $  & $ 0.4 - 1.5 $ \\ 
        PG 2349-014   & 357.9838 &  -1.1537 & 0.1738 & 15.4 & -23.5 &  0.05 & $ 10^{\prime\prime}$  &  S1.2  &  519  & $ 9.0 \times 10^{-6} - 4.2 \times 10^{6} $  & $ 0.9 - 1.0 $ \\ 
        PKS 1345+12   & 206.8892 &  12.2900 & 0.1205 & 17.1 & -22.0 &  0.03 & $ 10^{\prime\prime}$  &  S2    &  303  & $ 1.3 \times 10^{-4} - 8.7 \times 10^{6} $  & $ 1.0 - 3.9 $ \\ 
            Ton 951   & 131.9269 &  34.7512 & 0.0640 & 14.4 & -22.7 &  0.05 & $ 10^{\prime\prime}$  &  S1.0  &  251  & $ 8.6 \times 10^{-2} - 9.4 \times 10^{2} $  & $ 0.9 - 1.1 $ \\ 
              W Com   & 185.3820 &  28.2329 & 0.1020 & 15.3 & -23.1 &  0.03 & $ 10^{\prime\prime}$  &  BL    & -     & $ 8.2 \times 10^{-2} - 9.0 \times 10^{1} $  & $ 0.9 - 2.3 $ \\ 
 		\hline
	\end{tabular}
~\\   
$\dagger$ Classifications taken from \citet{veron}.    
\end{table*}

\begin{table*}
	\centering
	\caption{The sources of spectroscopy used for the SEDs.}
	\label{table:specsummary}
    \scriptsize
	\begin{tabular}{llllllll} 
		\hline
 Name & \multicolumn{7}{c}{Sources of spectra} \\
       & $<0.01~{\rm \mu m}$ & $0.09-0.33~{\rm \mu m}$ & $0.33-1.0~{\rm \mu m}$ & $1.0-2.5~{\rm \mu m}$  & $2.5-5~{\rm \mu m}$  & $5-42~{\rm \mu m}$ & $42-500~{\rm \mu m}$         \\
		\hline
2MASX J13000533+1632151   &                  &  9              &  9, 14          &  2              &  2              &  1              &                 \\ 
             3C 120   &  6              &  9, 12          &  4,  9          &  4,  9          &  2              &  1              &                 \\ 
             3C 273   &  6              &  7              &  3,  7,  9      &  3              &  2              &  1              &                 \\ 
             3C 351   &                  &  7              &  5,  7, 14      &  5              &  2              &  1              &                 \\ 
           3C 390.3   &  6              &  9, 10, 12      &  4,  9          &  4              &  2              &  1              &                 \\ 
            Ark 120   &  6              & 10, 11          &  3, 10          &  3              &  2              &  1              &                 \\ 
            Ark 564   &                  &  8, 10, 11      &  5, 10, 16      &  5              &                  &  1              &                 \\ 
          Fairall 9   &  6              & 12, 13, 18      & 18              &  2, 18          &  2              &  1              &                 \\ 
       F2M1113+1244   &                  &                  & 14, 39          &  2, 39          &  2              &  1              &                 \\ 
         H 1821+643   &                  & 10, 13          &  3, 27, 34      &  3              &  2              &  1              &                 \\ 
    IRAS 11119+3257   &                  &                  & 14              &  2              &  2              &  1              &                 \\ 
   IRAS F16156+0146   &                  & 18              & 18              & 18, 30          &  2              &  1              &                 \\ 
            Mrk 110   &  6              &  9, 11, 12      &  3,  9          &  3              &  2              &  1              &                 \\ 
           Mrk 1502   &                  & 10, 11          & 10, 31, 32, 38  & 38              &  2              &  1              &                 \\ 
            Mrk 231   &  6              &  8,  9, 10      & 36, 40          &  2, 40          &  2              &  1              & 19, 20, 21     \\ 
            Mrk 279   &  6              & 12, 13          &  5, 12, 41      &  5              &  2              &  1              &                 \\ 
            Mrk 290   &  6              &  7              &  3,  7          &  3              &  2              &  1              &                 \\ 
            Mrk 421   &  6              & 13, 22          & 37              &                  &                  &  1              &                 \\ 
            Mrk 493   &                  & 10              &  5, 10, 14      &  5              &                  &  1              &                 \\ 
            Mrk 509   &  6              &  7              &  3,  7          &  2,  3          &  2              &  1              &                 \\ 
            Mrk 590   &  6              &  8, 12          &  3, 14          &  3              &  2              &  1              &                 \\ 
            Mrk 817   &  6              &  8,  9, 11, 12  &  3, 41          &  3              &  2              &  1              & 24             \\ 
            Mrk 876   &  6              &  8, 11, 12      &  3, 12, 31      &  3              &  2              &  1              &                 \\ 
            Mrk 926   &  6              & 11, 12          & 14, 15, 27      & 27              &                  &  1              &                 \\ 
   NGC 3227 Central   &  6              &  9, 13          &  5,  9          &  5              &                  &  1              & 24             \\ 
   NGC 3516 Central   &  6              &  7,  9          &  4,  9, 27      &  4              &                  &  1              &                 \\ 
   NGC 4051 Central   &  6              &  8, 11, 12      &  5, 16          &  5              &                  &  1              &                 \\ 
   NGC 4151 Central   &  6              &  9, 11          &  5,  9          &  2,  5          &  2              &  1              &                 \\ 
   NGC 5548 Central   &  6              & 11, 13, 16      &  3, 14, 16      &  3              &  2              &  1              &                 \\ 
           NGC 5728   &  6              & 12              & 15, 28          & 23              &  2              &  1              &                 \\ 
           NGC 7469   &  6              & 10, 11          &  5, 10, 15, 29  &  5              &  2              &  1              & 19, 20, 21     \\ 
             OQ 530   &                  & 26              & 14              &                  &                  &  1              &                 \\ 
        PG 0026+129   &  6              & 10, 11, 17      &  4, 15, 17      &  2,  4          &  2              &  1              &                 \\ 
        PG 0052+251   &  6              &  7              &  4,  7          &  4              &  2              &  1              &                 \\ 
        PG 1211+143   &                  &  8, 12, 13, 31  &  4, 12, 31      &  4              &  2              &  1              &                 \\ 
        PG 1307+085   &                  &  8, 10, 11, 12, 31  &  4, 14, 31      &  4              &  2              &  1              &                 \\ 
        PG 1415+451   &                  & 10, 11          &  5, 14, 33      &  5              &  2              &  1              &                 \\ 
        PG 2349-014   &  6              &  7              &  7, 35          & 35              &                  &  1              &                 \\ 
        PKS 1345+12   &                  &                  & 14              &  2              &  2              &  1              &                 \\ 
            Ton 951   &                  &  7              &  3,  7          &  3              &  2              &  1              &                 \\ 
              W Com   &                  & 25              & 14              &                  &                  &  1              &                 \\ 
\hline
\end{tabular}
\\
1 {\it Spitzer} IRS, 2 {\it Akari}, 3 \citet{landt2008}, 4 \cite{landt2013}, 5 \cite{riffel2006}, 6 This work, 7 \cite{shang2005}, 8 HSLA \citep{peeples2017}, 9 STIS, 10 FOS, 11 FUSE, 12 IUE, 13 HUT, 14 SDSS, 15 MUSE, 16 AGN Watch, 17 UVES, 18 XShooter, 19 {\it Herschel} SPIRE, 20 {\it Herschel} PACS, 21 ISO, 22 WUPPE, 23 SINFONI, 24 MIPS SED, 25 COS, 26 GALEX, 27 BASS \citep{koss2017,lamperti2017}, 28 S7 \citep{dopita2015}, 29 \citet{kim1995}, 30 \citet{veilleux1999}, 31 \citet{wilkes1999}, 32 \citet{jansen2000}, 33 Modified \citet{kuraszkiewicz2000}, 34 \citet{decarli2008}, 35 \citet{glikman2006}, 36 \citet{rodriguez2009}, 37 \citet{smith2009}, 38 \citet{garcia2012}, 39 \citet{glikman2012}, 40 \citet{leighly2014}, 41 \citet{barth2015}
\end{table*}

\subsection{X-ray spectra}

To increase the utility of the SEDs, we have expanded the SEDs into the X-ray for the 27 objects that were detected by the 70-month catalogue \citep{Baumgartner:2013uq} of the all-sky survey carried out by the Burst Alert Telescope (BAT, \citealp{Barthelmy:2005uq}) on-board the {\it Neil Gehrels Swift Observatory} \citep{Gehrels:2004kx}. 

The bulk of our X-ray spectra come from \citet{ricci2017}, who undertook a detailed broad-band (0.3--150\,keV) X-ray spectral analysis of the 836 AGN from the {\it Swift}/BAT 70-month survey in the framework of the {\it BAT AGN Spectroscopic Survey\footnote{www.bass-survey.com}} \citep[BASS; ][]{berney2015,koss2017,lamperti2017}. This was done using a variety of models and X-ray spectra from {\it XMM-Newton} \citep{Jansen:2001ve}, {\it Chandra} \citep{Weisskopf:2000vn}, {\it Swift}/XRT \citep{Burrows:2005vn}, {\it Suzaku} \citep{Mitsuda:2007dq} and {\it Swift}/BAT. To homogenise the fluxes of the individual X-ray spectra and incorporate them into our SEDs, we have scaled the spectra to agree with the {\it Swift}/BAT flux, which was integrated over a period of 70 months, and could therefore be considered as a good indicator of the mean AGN emission. We caution that by scaling to the mean X-ray flux, we do reduce the impact of variability on our SEDs, and single epoch observations of AGNs may show more X-ray variabilty (e.g. a broader spread of X-ray to IR flux ratios) than our SEDs. The ratios between the flux measured by 0.3--10\,keV observations and the time-averaged {\it Swift}/BAT flux for this sample are found to be in the $\sim 0.4-4.3$ range. 

For 16 objects we also include additional X-ray spectra from {\it NuSTAR} \citep{Harrison:2013zr}, typically in the 3--40\,keV range, with the {\it NuSTAR} spectra also being scaled to the {\it Swift}/BAT flux produce continuous X-ray spectra. To maximise the signal-to-noise ratio we used the following approach whenever there was overlapping spectral coverage from multiple missions. i) For sources with {\it NuSTAR}, {\it XMM-Newton} or {\it Suzaku} observations we used {\it XMM-Newton} ({\it Suzaku}) observations in the 0.3--10\,keV (0.5--10\,keV) range, {\it NuSTAR} above 10\,keV, up to the energy range in which the background radiation would dominate ($E_{\rm bkg}$). Between $E_{\rm bkg}$ and 150\,keV we used {\it Swift}/BAT data. ii) For sources with {\it NuSTAR} and {\it Swift}/XRT observations we used {\it Swift}/XRT in the 0.3--3\,keV interval, {\it NuSTAR} between 3\,keV and $E_{\rm bkg}$, and {\it Swift}/BAT in the $E_{\rm bkg}$--150\,keV range.

In several instances the {\it NuSTAR} spectra clearly show the narrow iron K$\alpha$ lines (e.g., \citealp{Shu:2010qv}), including the spectra of Mrk~590, Mrk~926, NGC~3227 and NGC~5728. The sample also includes very diverse X-ray spectral continuum shapes, with several objects showing ionised absorbers (e.g., NGC\,4051, 3C\,351; see \citealp{Tombesi:2013yo} and references therein), and in one case (NGC\,5728, \citealp{Ricci:2015kx}) Compton-thick obscuration ($N_{\rm\,H}\geq 10^{24}\rm\,cm^{-2}$). 

\subsection{Modelling far-infrared and radio spectra}

Relatively few AGNs in our sample have far-infrared spectrophotometry (see Table~\ref{table:specsummary}) and we do have not have any radio spectra with broad frequency coverage. However, as the far-infrared is often dominated by thermal emission from warm dust and radio SEDs are approximately power-laws, we can produce functional SEDs by fitting models to the observed fluxes. 

We model the thermal emission from warm dust by fitting greybody curves to published far-infrared photometry (when available) from the Herschel/PACS Point Source Catalogue \citep[HPPSC; ][]{marton2017}, the Herschel/SPIRE Point Source Catalogue \citep[HSPSC; ][]{schulz2017}, the WMAP point source catalogue \citep[][]{bennett2013} and the {\it Planck} point source catalogue \citep[][]{planck2016}. As we are fitting six (or fewer) photometric data points with a model with four free parameters, some caution is required and all fits to the data are visually inspected. 

We have been able to extend some of the SEDs to radio wavelengths using published radio flux density measurements to constrain simple empirical models. The radio portion of the SEDs is not as precise as the SEDs at shorter wavelengths, but will still be useful for some purposes (e.g. modelling the range of infrared flux densities of radio selected quasars). Radio flux density measurements are drawn from the WMAP point source catalogue \citep[][]{bennett2013}, the {\it Planck} point source catalogue \citep[][]{planck2016}, GLEAM \citep{hur16}, the compilation of \citet{kuehr1981} and the NASA Extragalactic Database (NED). While the WMAP, {\it Planck} and GLEAM flux density measurements are relatively homogeneous, most of the radio flux density measurements come from a variety of sources and some caution is required when modelling the data. 

To model radio spectra, we fitted polynomials to log radio flux density as a function log frequency. For 3C~351 and PKS~1345+12 we have used the previously published models of \citet{kuehr1981} for flux density as a function of wavelength. As the input radio fluxes are inhomogenous we cross checked that we were using (effectively) total flux densities and compared the model fits to the measured flux densities (to search for discontinuities resulting from variability etcetera). In some instances a tight relation between radio flux density and frequency is not evident and we couldn't model the radio SED. However, in other instances the polynomials are good approximations of the observed flux densities (scatter of $\sim 0.1~{\rm dex}$) so we include these models in our AGN SEDs. 

\subsection{Photometry}

To scale individual spectra and cross check the validity of the SEDs, we use matched aperture photometry measured from images taken with GALEX, {\it Swift}, SDSS, PanSTARRS, Skymapper, 2MASS, WISE and {\it Spitzer}. The photometry was measured using circular apertures with the diameters given in Table~\ref{table:agnsummary}. Point spread function corrections are applied to the GALEX, WISE and {\it Spitzer} photometry, and for the longer wavelengths we also expand the aperture size so the point spread function corrections never exceed a factor of two. For the {\it Swift} photometry we apply point source coincidence loss corrections, while cautioning that these will be inaccurate for extended sources \citep[e.g.][]{bro14}. All photometry (and relevant spectra) are corrected for Milky Way foreground dust extinction using the {\it Planck} dust maps \citep{pla11,pla13} and a \citet{fit99} model.

All $0.1$ to $30~{\rm \mu m}$ photometry was measured using the AB magnitude system and the corresponding flux densities were determined using $f_\nu \simeq 3631~{\rm Jy} \times 10^{-0.4m}$ where $m$ is the apparent magnitude. Consequently our photometry may systematically differ from the default photometry in some bands (e.g. when the default zeropoints are defined using a $f_\nu \propto \nu^{2}$ source). The WISE $W4$ band effective wavelength differs from what was measured prior to launch, and we have corrected for this using the methodology described in \citet{bro14b}. Our photometry is summarised in Table~\ref{table:obsfluxes}.

\section{Constructing SEDs}
\label{sec:construction}

\subsection{Combining spectra}

Our SEDs are anchored in the publicly available spectra summarised in Table~\ref{table:specsummary}. In instances where multiple individual extracted spectra are available, as is often the case for IUE and Akari, we combine the individual spectra together to achieve higher signal-to-noise. The individual spectra are also visually inspected, and we mask strong geocoronal features and atmospheric absorption features.

While we mostly use publicly available reduced spectra in the $0.09$ to $500~{\rm \mu m}$ range, we reprocessed archival ESO XShooter and SINFONI spectra. We reduced archival XShooter spectra\footnote{http://archive.eso.org/} using version 2.9.3 of the XShooter pipeline \citep{modigliani2010} accessed via the Reflex interface, followed by telluric line corrections with molecfit \citep{smette2015}. Mark Durr\'{e} kindly reduced SINFONI spectra of NGC~5728 as described in \citet{durre2018}, using the SINFONI data reduction pipeline (version 2.5.2) accessed through the gasgano interface, with telluric correction and flux calibration from standard stars observed at a similar airmass.

The variability of AGNs, exemplified by changing-look quasars,  complicates the production of continuous spectra of individual objects. We are utilising spectra and photometry taken over years (and in some instances decades) and inevitably archival data will produce inconsistent flux densities at some wavelengths. To produce continuous SEDs we take the input spectra and apply multiplicative scalings to mitigate discontinuities resulting from the variety of spectrographs, extraction apertures and AGN variability. Preliminary scalings are estimated by comparing spectra with photometry, but AGN variability and aperture bias results in this producing discontinuities, particularly in the ultraviolet and near-infrared. When input spectra overlap in wavelength, we measure the ratio of the flux densities in the relevant wavelength range, and use this to determine scalings that produce continuous spectra. However, as input spectra can have relatively large random and systematic errors at the extrema of their wavelength ranges, manual inspection and intervention is required to produce continuous spectra. In some instances the shape of input spectra taken at different wavelengths (or by different instruments) is inconsistent and does not approximate a plausible SED (even after rescaling), and the relevant AGN has to be rejected from the sample (e.g., Mrk~3, NGC~1068, NGC~6814, PG~0804+761, PG~1302-102). 

\begin{figure}
 \includegraphics[width=\columnwidth]{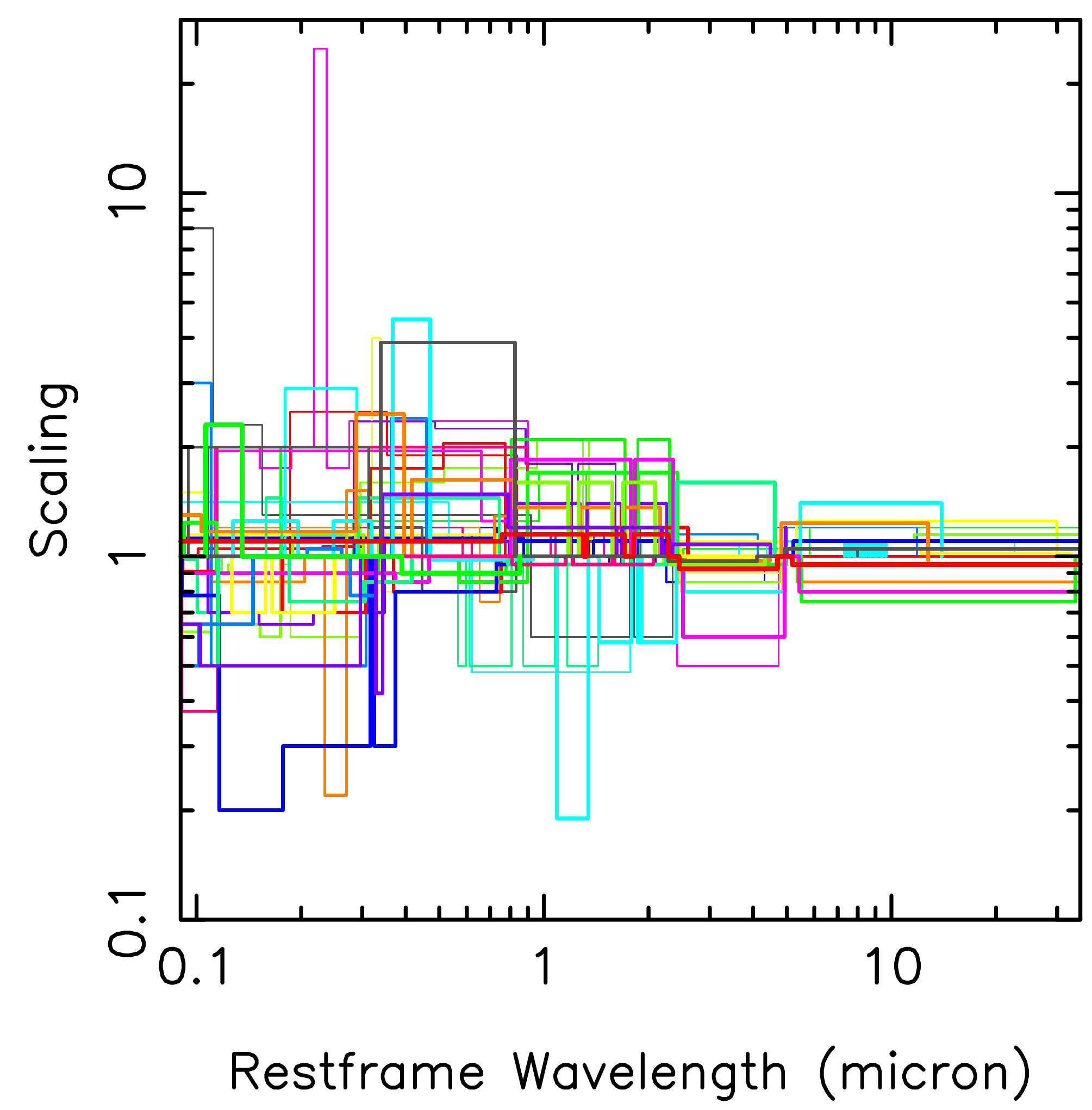}
  \caption{The multiplicative scalings applied to the individual spectra used to produce the AGN SEDs. Variability and aperture bias results in the range of scalings increasing as one moves to shorter wavelengths. For half the sample the scalings never exceed the 0.5 to 2.0 range, but 7 AGNs have scalings that fall outside the 0.33 to 3.0 range (3C 390.3, Ark 120, Mrk 110, Mrk 509, NGC 5548, NGC 5728 and PKS 1345+12).}
 \label{fig:scaling}
\end{figure}

The multiplicative scalings applied to the individual spectra are plotted in Figure~\ref{fig:scaling} as a function of wavelength and the range of scalings applied to each object is provided in Table~\ref{table:agnsummary}. For some of our SEDs the multiplicative scalings alter the input flux densities by tens of percent, but in some instances input and final flux densities differ by more than a factor of 3. As AGN variability increases with decreasing wavelength, the scalings plotted in Figure~\ref{fig:scaling} show more scatter in the ultraviolet than in the mid-infrared. We caution that multiplicative scaling of individual input spectra is a crude approximation, and the variability and aperture bias will have some wavelength dependence within the individual input spectra (e.g., emission lines relative to continuum). For some purposes, users of our SEDs may want to restrict themselves to AGN SEDs with a relatively limited range of multiplicative scalings and review papers describing the sources of the individual spectra (which are provided in Table~\ref{table:specsummary}).

All of the SEDs have gaps in their spectral coverage, including gaps resulting from atmospheric absorption in the near-infrared, gaps in spectral coverage (i.e., between {\it Akari} and {\it Spitzer}) and gaps between data and models (i.e., {\it Spitzer} and the far-infrared). To interpolate across these gaps we fit polynominals to portions of the spectra on either end of the relevant gap (fitting to log flux density as a function of log wavelength). In some instances we cannot fill (large) gaps in spectral coverage (e.g., X-ray and ultraviolet, far-infrared and radio) so we leave these gaps in the SEDs. When ultraviolet spectra are available but they do not extend down to $0.09~{\rm \mu m}$, we extrapolate the ultraviolet using a power-law (fitted to log flux density as a function of log wavelength). 

Several obscured quasars, Seyferts and BL Lacs that lack ultraviolet and/or {\it Akari} spectra are included in the sample (which would otherwise be completely dominated by broadline quasars). As the {\it Akari} spectra are relatively simple (i.e., lacking strong emission lines), we have interpolated across the $2.4$ to $5.5~{\rm \mu m}$ wavelength range using polynomials. As BL Lacs are notable for their lack of spectral features and their spectra are almost power-laws, we have used polynomials to interpolate and extrapolate their spectra over large portions of the $0.09$ to $5.5~{\rm \mu m}$ wavelength range. 

Obscured quasars often show an ultraviolet excess relative to extrapolations of their optical SEDs (i.e., Mrk~231). Consequently, we extrapolate their spectra into a ultraviolet using a toy model, which includes continuum short and long of Lyman-$\alpha$, plus the Lyman-$\alpha$ line itself. The model is deliberately crude so it cannot be confused with data or a physically motivated model, with $f_\lambda$ fixed at constant values above and below Lyman-$\alpha$, with Lyman-$\alpha$ emission approximated by a step function. Despite its simplicity, this toy model allows the obscured quasar SEDs to be used when restframe ultraviolet wavelength is required, such as $z>2$ photometric redshifts (although the restframe ultraviolet may be practically undetectable). 

\begin{figure}
 \includegraphics[width=\columnwidth]{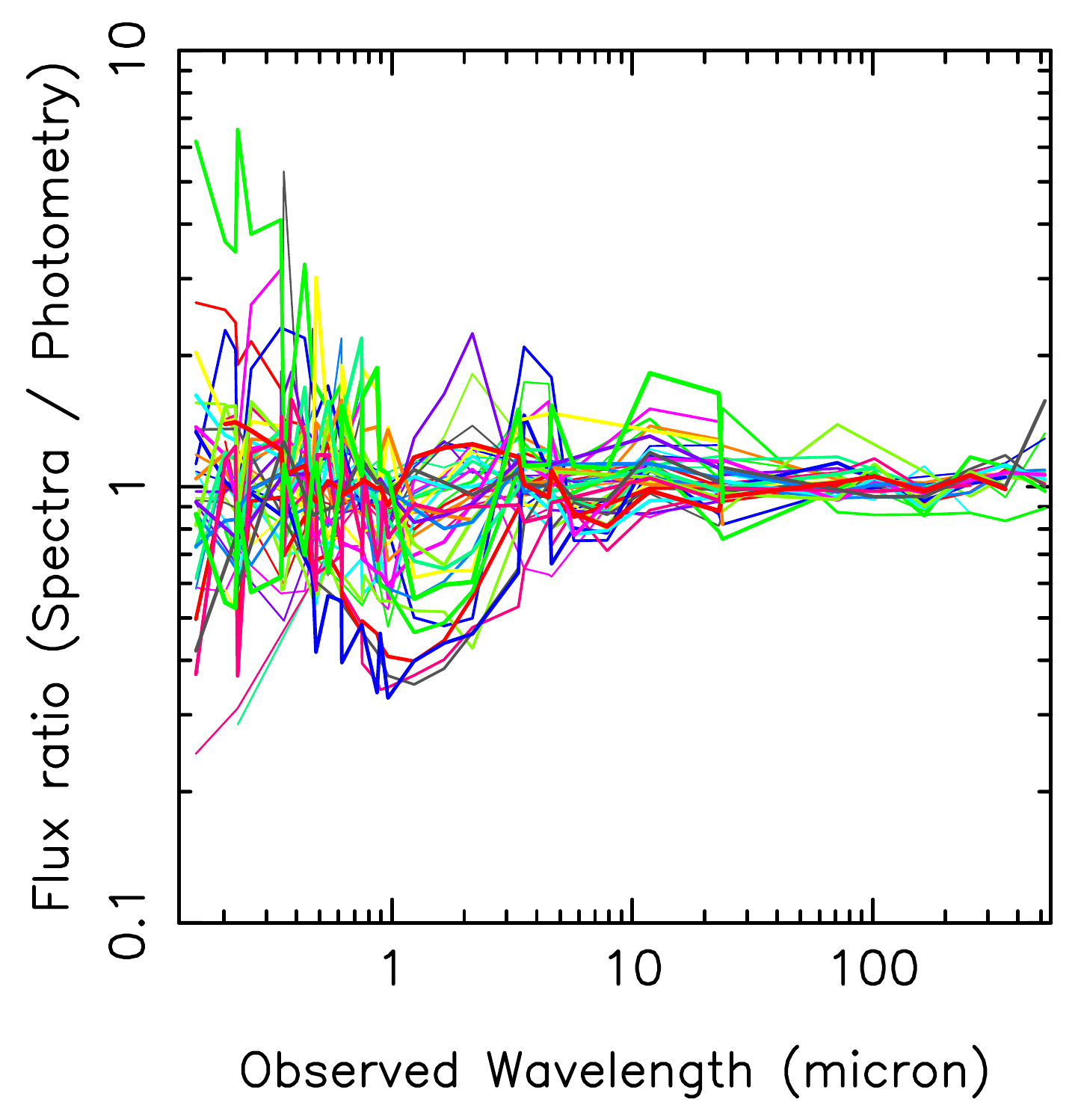}
  \caption{A comparison of measured aperture photometry and photometry synthesised from the SEDs. The photometry includes more host galaxy light than the SEDs, resulting in systematic offsets at $1~{\rm \mu m}$, while variability results in increased scatter with decreasing wavelength.}
 \label{fig:fluxratio}
\end{figure}

In Figure~\ref{fig:fluxratio} we present a comparison of fluxes synthesised from the 41 SEDs with fluxes measured directly from images. As with the scalings used for the spectra, the difference between synthesised and measured photometry generally increases with decreasing wavelength and there's an offset at $\sim 1~{\rm \mu m}$ resulting from the aperture photometry including host galaxy light\footnote{A curious exception to the trend of variabilty increasing with decreasing wavelength is Mrk~926, where the 2MASS $JHK$ photometry and {\it Spitzer} spectroscopy at $\sim 6~{\rm \mu m}$ falls below more recent $JHK$ spectroscopy, {\it Spitzer} photometry and WISE photometry. Mrk 926 also features asymmetric and variable broad emission lines \citep[i.e.][]{kollatschny2010}.}. Differences between synthesised and measured fluxes are not a smooth function of wavelength, as we sometimes have multiple photometric measurements taken at similar wavelengths but at different epochs. For example, SDSS, Skymapper and PanSTARRS provide $griz$ photometry while {\it Spitzer} and WISE have three bands with comparable effective wavelengths. The flux densities synthesised from the SEDs are provided in Table~\ref{tab:synfluxes}, and these have less variation as a function of wavelength than the directly measured photometry.

In Figure~\ref{fig:iauqsosed} we provide an illustrative $0.09$ to $600~{\rm \mu m}$ SED of a broadline quasar, PG~0052+251, along with illustrative quasar SEDs from the past three decades. The prior literature illustrates the increasing availability of data with time, different methodologies, and compromises one has to make between spectral coverage, spectral resolution and wavelength range. Relative to the prior literature, we have greater spectroscopic coverage, which results in more emission lines and mid-IR spectral features that were largely absent from previous SEDs. The shape of our PG~0052+251 SED in the mid and far-IR is significantly different from that of most previous quasar templates, but we caution that quasar SEDs show significant diversity at these wavelengths \citep[e.g.,][]{netzer2007,lyu2017a,lyu2017b}, a point that we shall return to later in this paper. 

\begin{figure*}
\includegraphics[width=\textwidth]{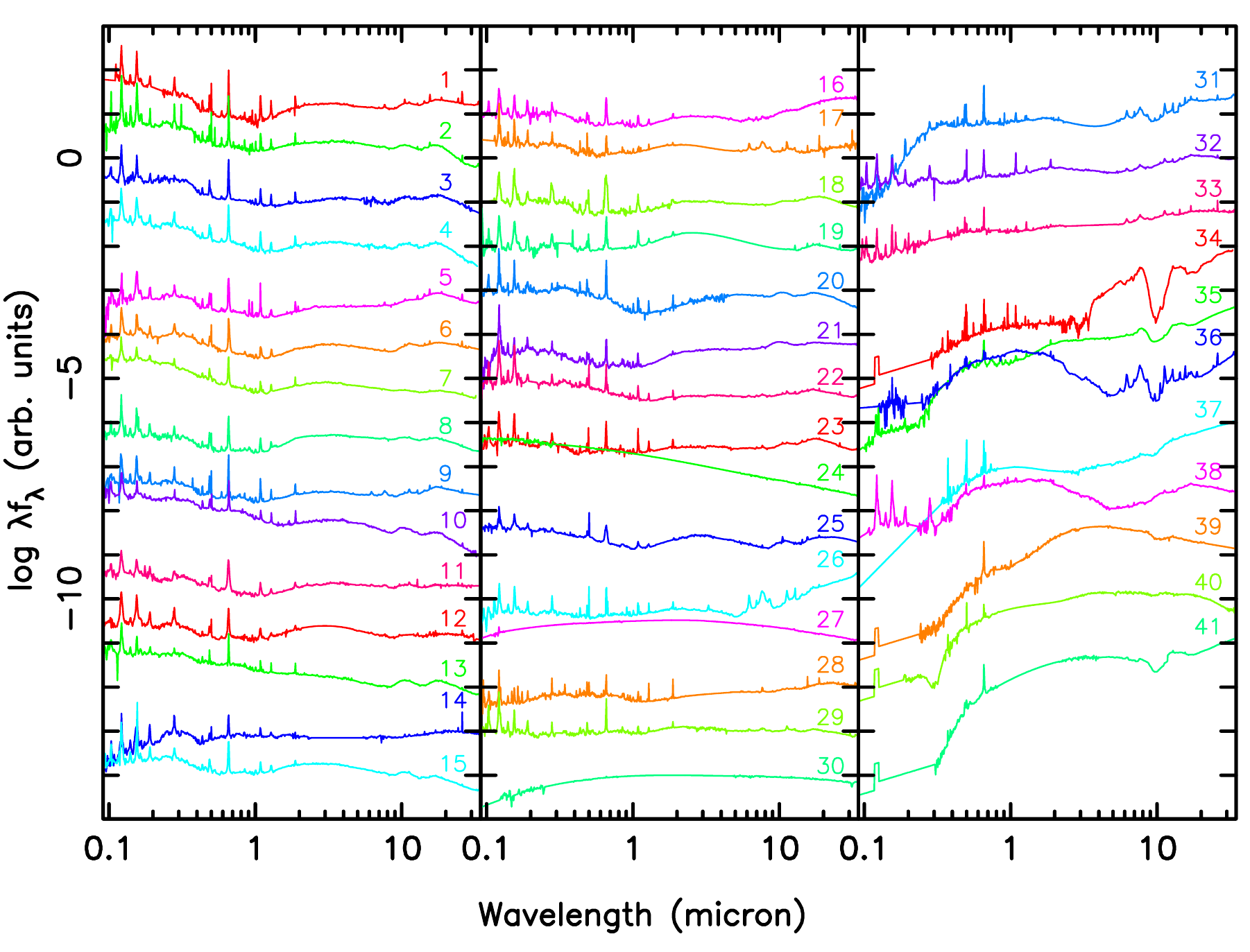}
  \caption{The SEDs of the 41 individual AGNs sorted by restframe $u-g$ colour, with the bluest templates at the top-left of the plot. UV and optical emission lines are immediately evident. The $9.7~{\rm \mu m}$ silicate absorption is evident for some red AGN templates while $9.7~{\rm \mu m}$ silicate emission can be discerned for some blue quasars. The numbers on the individual SEDs correspond to: 1. 3C 120; 2. Mrk 110; 3. PG 1307+085; 4. PG 0052+251; 5. NGC 4151 Central; 6. H 1821+643; 7. 3C 273; 8. Fairall 9; 9. Mrk 509; 10. PG 0026+129; 11. Mrk 876; 12. PG 2349-014; 13. Ton 951; 14. NGC 3516 Central; 15. Ark 120; 16. Mrk 817; 17. Mrk 493; 18. 3C 390.3; 19. Mrk 926; 20. PG 1211+143; 21. Mrk 1502; 22. Mrk 279; 23. Mrk 290; 24. Mrk 421; 25. 3C 351; 26. NGC 7469; 27. W Com; 28. Ark 564; 29. PG 1415+451; 30. OQ 530; 31. NGC 3227 Central; 32. NGC 5548 Central; 33. NGC 4051 Central; 34. IRAS F16156+0146; 35. Mrk 231; 36, NGC 5728; 37. PKS 1345+12; 38. Mrk 590; 39. F2M1113+1244; 40. 2MASX J13000533+1632151; 41. IRAS 11119+3257.}
\label{fig:allseds}
\vspace*{8cm}
\end{figure*}

In Figure~\ref{fig:allseds} we present the $0.09$ to $35~{\rm \mu m}$ SEDs of the 41 individual AGNs, with the SEDs ordered by $u-g$ colour. The similarity of the blue quasar SEDs to each other is apparent, although one can also see differences in the mid-infrared spectra at $\sim 20~{\rm \mu m}$. Figure~\ref{fig:allseds} illustrates that we have a relatively limited number of red AGN SEDs, and that our red AGN SEDs often utilise toy models and/or power-law interpolation. The red AGN population also shows considerable diversity, including the strength of ${\rm H\alpha}$ emission and $9.7~{\rm \mu m}$ silicate absorption. 

\begin{figure*}
 \includegraphics[width=\textwidth]{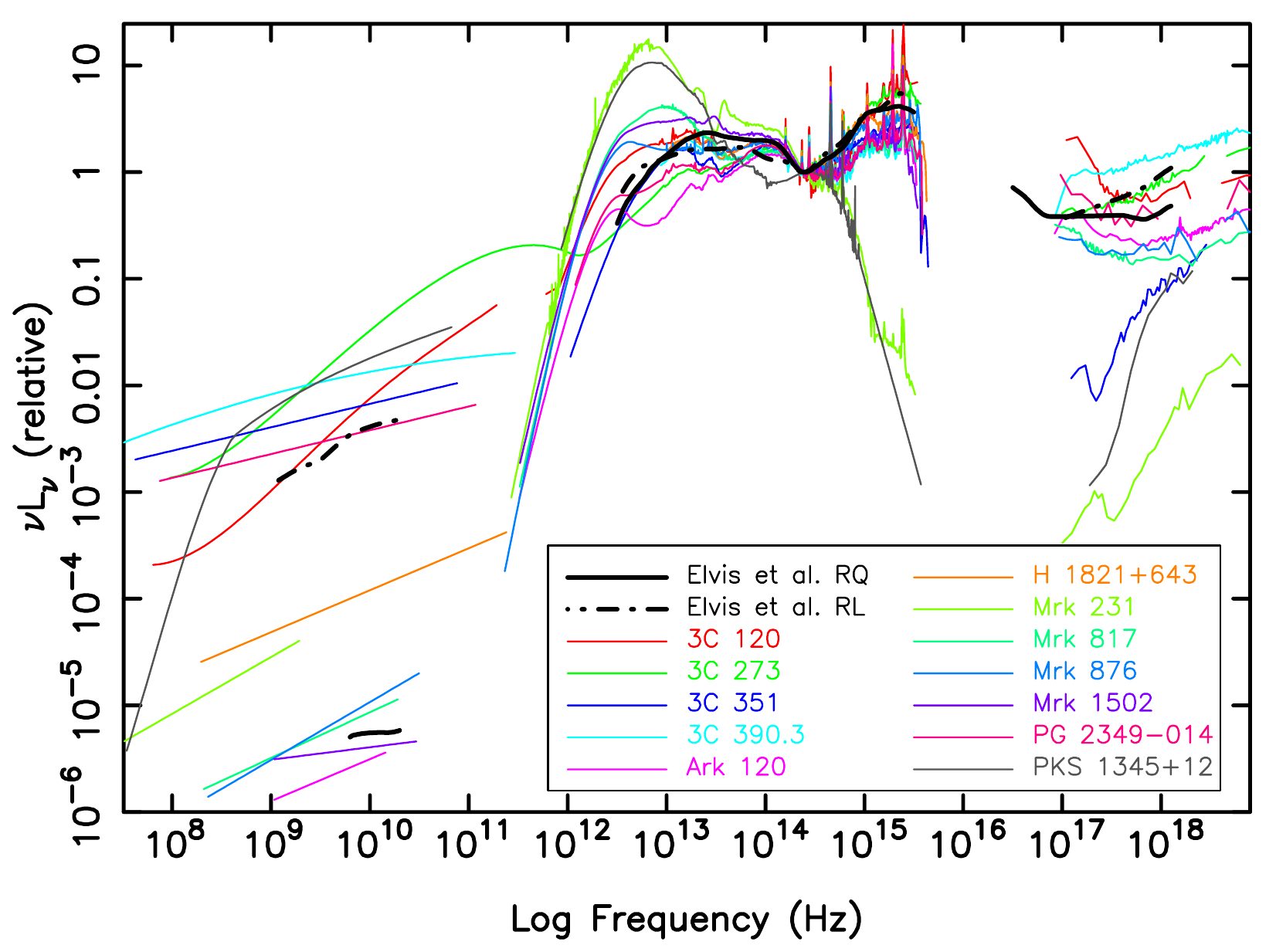}
  \caption{The SEDs of AGNs that have radio coverage, normalised at $1.25~{\rm \mu m}$. For comparison we also plot the median radio-loud and radio-quiet SEDs of \citet{elvis1994}. Several features that are evident, including the change of spectral index for 3C~120, 3C~273 and PKS~1345+12 at low frequencies, the diversity of infrared SEDs, and most of our SEDs being redder in the ultraviolet-optical than those of \citet{elvis1994}.}
 \label{fig:elvis}
\vspace*{8cm}
\end{figure*}

In Figure~\ref{fig:elvis} we plot the radio to X-ray SEDs for AGNs that have radio coverage. To facilitate comparison of the SEDs, we have normalised them at $1.25~{\rm \mu m}$ and overplotted the median radio-loud and radio-quiet SEDs from \citet{elvis1994}, which have been a useful benchmark over the past quarter of a century. Our blue SEDs are similar to those of  \citet{elvis1994}, including the factor of $\sim 1000$ difference in radio luminosity for radio-loud and radio-quiet quasars. That said, our SEDs are mostly redder than those of \citet{elvis1994} in the ultraviolet and optical, and in the far-infrared greybody curves can be steeper than the Rayleigh-Jeans approximation. While many of the radio SEDs are approximately power-laws, 3C~120, 3C~273 and PKS~1345+12 show changes in spectral index at low frequencies \citep[which has been previously noted by][]{white2018}.

\subsection{AGN and host galaxy composites}

Our sample is biased towards luminous quasars and the centres of nearby Seyferts where the observed spectrum is dominated by the AGN. We thus do not have many Seyfert galaxies where the fraction of AGN and host galaxy light varies greatly with wavelength, with host galaxy light peaking in the restframe optical and near-infrared. Seyferts comprise the bulk of AGNs detected by many surveys (particularly X-ray selected surveys at low redshift) so the absence of Seyfert templates could hinder the utility of our library, particularly for photometric redshifts \citep[e.g.,][]{salvato2009,hsu2014,salvato2018}, which is one of the principal motivations for this work.

We have thus created a series of AGN and host galaxy composites to produce Seyfert SEDs. To do this we have taken the AGN SEDs of NGC 3227, 3516, 4051 and 4151 and combined them with the galaxy SEDs of NGC 3310, 4125, 4138, 4569, 4725 and 4826 taken from \citet{bro14}. By construction this will produce SEDs that best mimic low redshift AGNs, and a different set of host galaxy SEDs could (or should) be used for Seyferts at higher redshift. As the wavelength coverage and precision of high redshift galaxy SEDs is poor relative to the \citet{bro14} SEDs, we do not create SEDs tailored specifically for high redshift Seyferts for this work. We combine the AGN and $z \sim 0$ host galaxy SEDs by renormalising the spectra at $0.6~{\rm \mu m}$ and then summing the SEDs using ratios of AGN to host light spanning from 2:1 to 1:64. This results in 72 additional AGN templates. 

\section{The colours of AGNs}
\label{sec:colours}

\begin{figure}
\centering
 \includegraphics[width=\columnwidth]{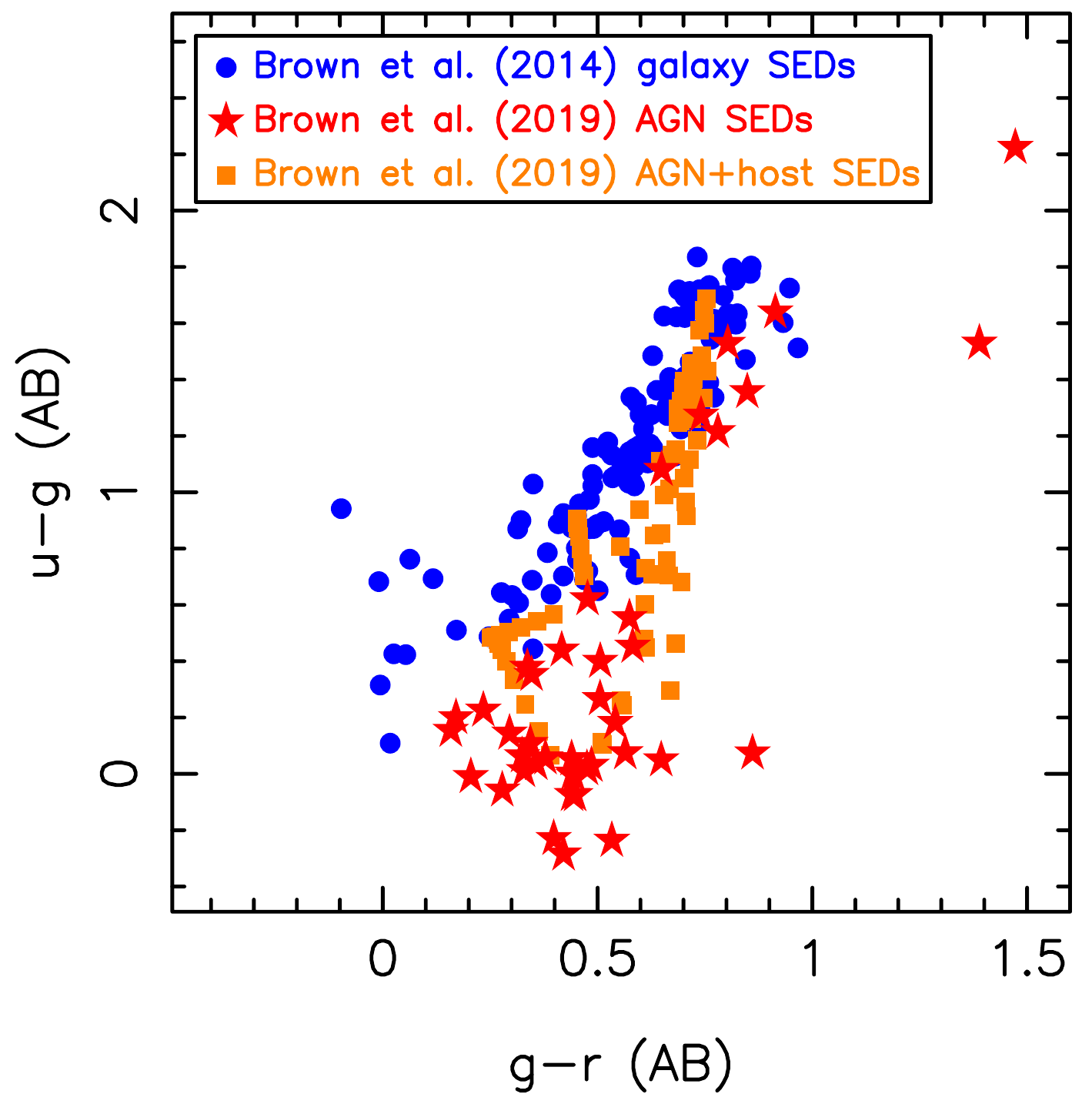}
  \caption{The restframe optical colours of the AGN SED templates (including AGN and host combinations) along with the colours of galaxy SED templates from \citet{bro14}. Blue continuum and strong ${\rm H\alpha}$ emission offset templates down and to the right (respectively) from other galaxies.}
 \label{fig:ugr}
\end{figure}

The synthesised colours of SED templates provide a useful summary of template library properties, including the diversity of the SEDs and differences relative to various object classes and the prior literature. In Figure~\ref{fig:ugr} we present the restframe $ugr$ optical colours of our AGN templates and the local galaxy templates of \citet{bro14}. \citet{bro14} used optical scanned spectra, {\it Akari} $2.5-4.9~{\rm \mu m}$ spectra and {\it Spitzer} $5.5-35~{\rm \mu m}$ spectra in combination with MAGPHYS models \citep{dac08} to produce SEDs for $z<0.04$ galaxies, and \citet{bro14} excluded many powerful AGNs due to the limitations of the version of MAGPHYS used at the time. Bright blue quasars are bluer in $u-g$ than galaxies, but are redder than galaxies in $g-r$ due to the contribution of ${\rm H\alpha}$ to the restframe $r$-band (the ${\rm H\alpha}$ equivalent widths are summarised in Table~\ref{table:agnsummary}). An extreme example of this is 3C 390.3, which has restframe colours of $u-g= 0.1$ and $g-r=0.9$. Reddened AGNs are also offset from the main locus of galaxies, including Mrk~231, which has restframe colours of $u-g=1.2$ and $g-r=0.8$. Extreme examples of reddened AGNs, including F2M~1113+1244 and IRAS~11119+3257, have optical colours unlike those of most red galaxies, including most LIRGS, with restframe $g-r>1$.

\begin{figure}
\centering
 \includegraphics[width=\columnwidth]{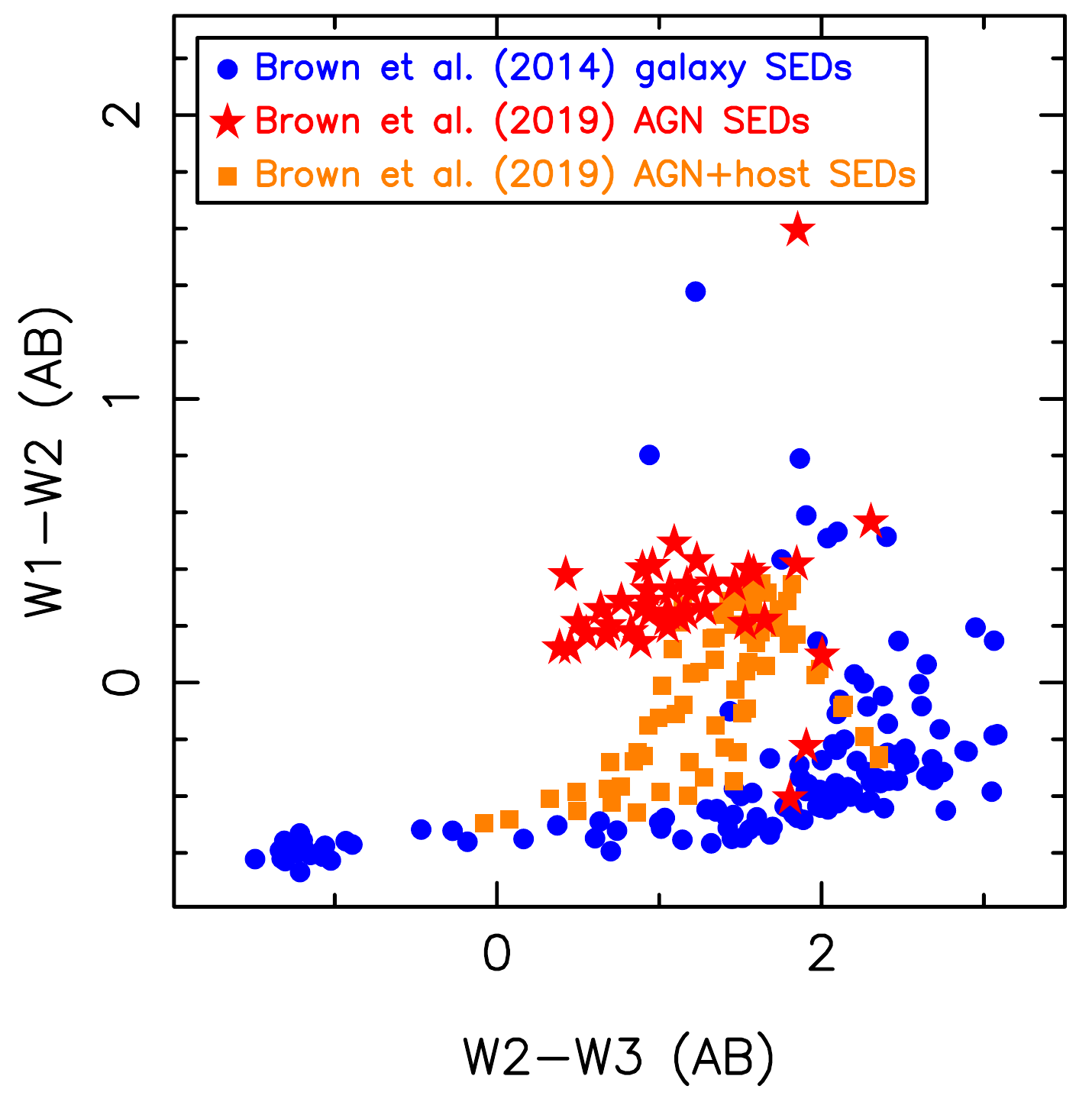}
  \caption{The restframe WISE mid-infrared colours of the AGN SED templates (including AGN and host combinations) along with the colours of galaxy SED templates from \citet{bro14}. The AGN templates are clustered in colour-space, and this is exploited by mid-infrared AGN selection criteria \citep[e.g.,][]{lacy2004,stern2005,assef2013}. AGN and host galaxy combinations form ``octopus legs'' that branch down towards the bulk of the \citet{bro14} galaxy templates.}
 \label{fig:wise}
\end{figure}

In Figure~\ref{fig:wise} we plot the restframe WISE mid-infrared colours of our AGN templates and the galaxy templates of \citet{bro14}. There is a locus of quasars that has systematically redder $W1-W2$ colours than most galaxies, and this locus is exploited by a number of mid-infrared quasar selection techniques \citep[e.g.,][]{lacy2004,stern2005,assef2013}. IRAS F16156+0146 and PKS~1345+12 both have mid-infrared colours significantly redder than most quasars, although in addition to being AGNs these two objects are also ULIRGs. The AGN-host galaxy combinations produce a series of ``octopus legs'' in Figure~\ref{fig:wise}, connecting the quasar locus with different populations of galaxies.  

The observed colours of SEDs as a function of redshift illustrate how they fill the colour-space and can reveal how individual spectral features impact broadband colours. To benchmark our work, we compare the colours produced by our SEDs to the \citet{bro14} galaxy SEDs, which largely lack powerful AGNs, and the \citet{ananna2017} galaxy and AGN SEDs, which are a recent iteration of AGN SEDs used for the SWIRE, CDF-S and COSMOS surveys \citep{pol07,salvato2009,hsu2014,ananna2017}. The \citet{ananna2017} suite of SEDs is also optimised for the wide-field Stripe 82 region, and thus may be directly comparable to our work (i.e. including luminous quasar SEDs).

\begin{figure}
\centering
 \includegraphics[width=\columnwidth]{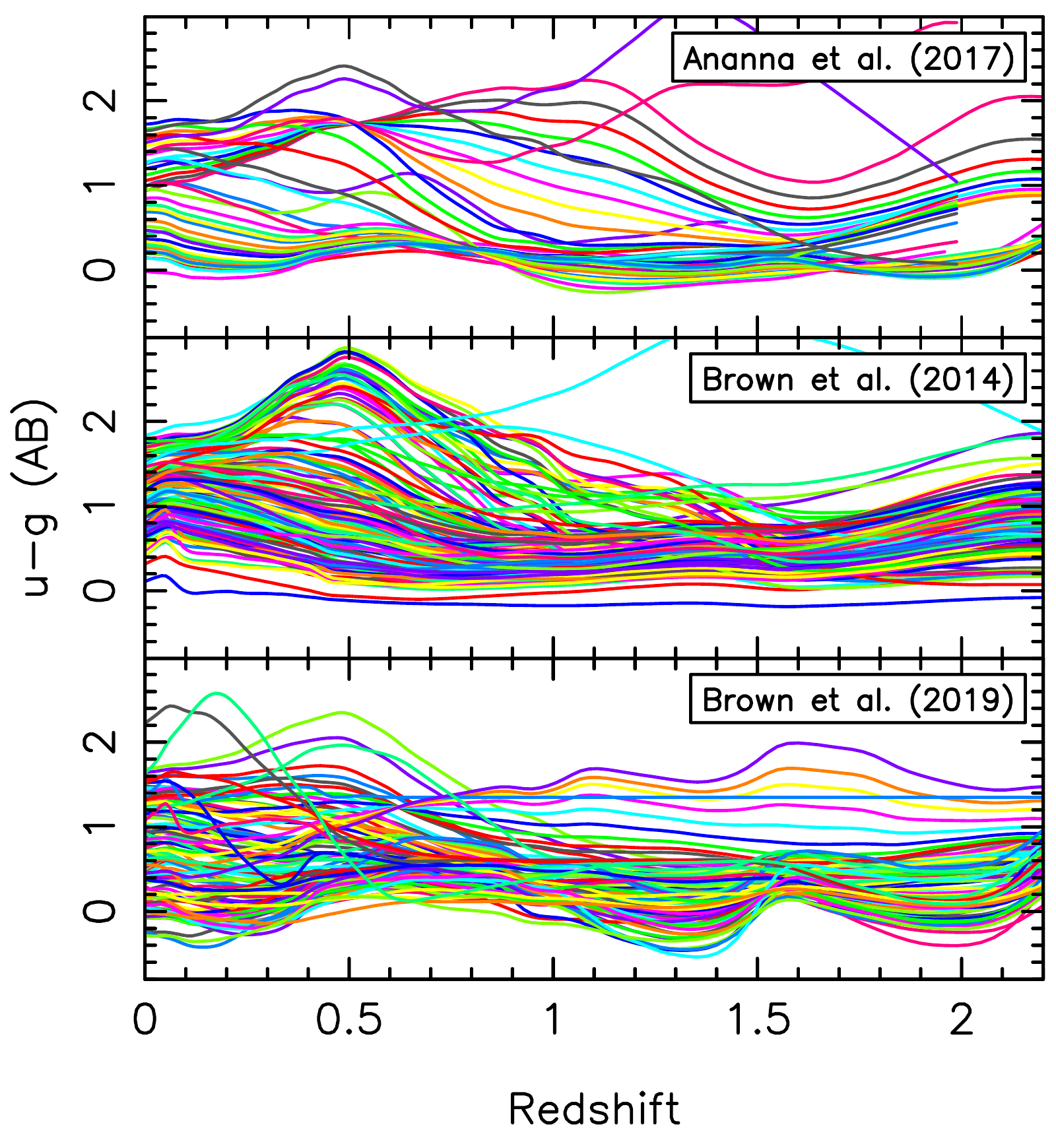}
  \caption{Observed $u-g$ colour as a function of redshift for the \citet{ananna2017} AGN and galaxy templates, the \citet{bro14} galaxy templates and the AGN templates from this work (including AGN and host combinations). The most notable difference between our templates and the others plotted above are our very blue quasar templates, including contributions by ultraviolet emission lines.}
 \label{fig:ug_redshift}
\end{figure}

In Figure~\ref{fig:ug_redshift} we plot observed $u-g$ colour as a function of redshift for the \citet{ananna2017} galaxy and AGN SEDs, \citet{bro14} galaxy SEDs and the AGN SEDs from this study. There are some obvious similarities between the different template libraries, including a red peak at $z\sim 0.5$ resulting from galaxy (or host galaxy) light. There are  clear differences between the template libraries too, including the very blue colours of some of our quasar templates and the contribution of very strong restframe UV emission lines (particularly in 3C and PG quasars) to the $u-g$ colours at $z\sim 1.4$ and $z\sim 2$. The variation of $u-g$ (and $U-B$) quasar colour as a function redshift is well established and produced by some template libraries \citep[e.g.,][]{vandenberk01}, it impacts  UV-excess quasar selection, and can be exploited for photometric redshifts \citep[e.g.,][]{weinstein2004}

\begin{figure}
\centering
 \includegraphics[width=\columnwidth]{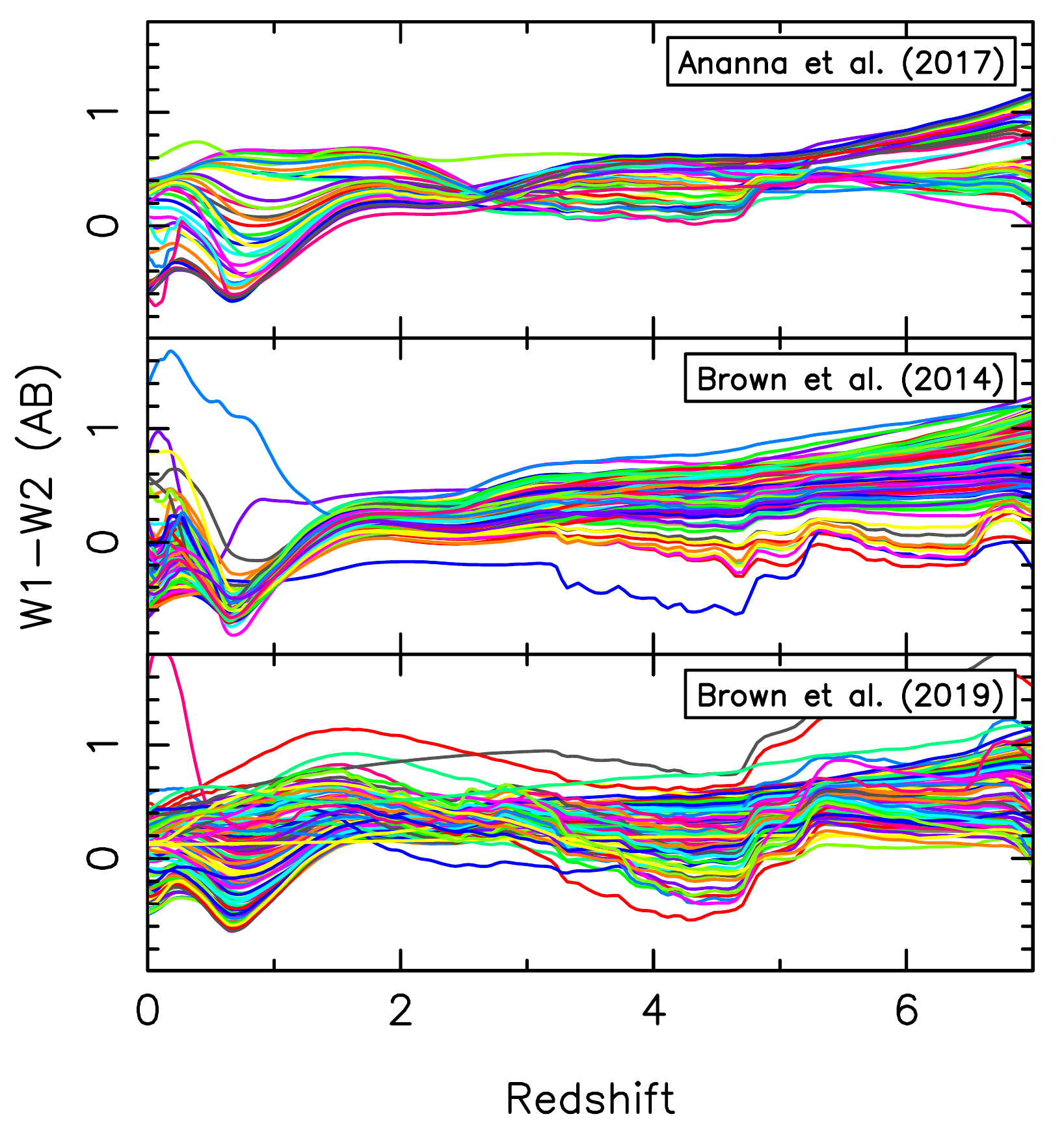}
  \caption{Observed WISE $W1-W2$ colour as a function of redshift for the \citet{ananna2017} AGN SED templates, the \citet{bro14} galaxy templates and the AGN templates from this work. One clear difference between the mid-infrared colours of our templates and some of those from the literature is evident at $z\sim 4$, where ${\rm H\alpha}$ is in the WISE $W1$ band.}
 \label{fig:wise_redshift}
\end{figure}

In Figure~\ref{fig:wise_redshift} we plot WISE infrared colours as a function of redshift for the \citet{ananna2017} SEDs, \citet{bro14} galaxy SEDs, and the AGN SEDs from this study. Low redshift quasars have significantly redder colours than galaxies, and lack strong contributions from the $3.3~{\rm \mu m}$ PAH feature and the $1.6~{\rm \mu m}$ ${\rm H^-}$ spectral feature. We find the WISE $W1-W2$ colours of quasars become gradually bluer with increasing redshift between $z\sim 1.5$ and $z\sim 3$, whereas the colours of galaxies (including some Seyferts) become redder over the same redshift range. At $z\sim 4$ the contribution of ${\rm H\alpha}$ to the WISE $W1$ flux results in very blue $W1-W2$ colours, and is thus a potentially useful feature for photometric redshifts.  

\section{Photometric Redshifts}
\label{sec:photoz}

Photometric redshifts (photo-$z$s) are one of the principal motivators for developing our AGN SED library and thus are a natural test of the SEDs validity and utility. Photo-$z$s work particularly well when the objects of interest have a limited range of restframe colours while having observed colours that are a strong function of redshift (exemplified by elliptical galaxies). While UV-excess selected quasars have a limited range of rest-frame optical colours, their mid-infrared SEDs show significant diversity and their colours are a relatively weak function of redshift at $z<2.3$. Thus quasar photo-$z$s have often been a challenge \citep[see the review of][ and references therein]{salvato2018}.

We test the utility our SEDs for photometric redshifts with AGNs selected from the Bo\"{o}tes field, which we and others have previously used for the development and testing of AGN photometric redshifts \citep[e.g.,][]{brodwin2006,duncan2018a,duncan2018b}. We utilise the matched aperture photometry initially described by \citet{bro07}, but then subsequently updated to now include LBT $uy$, {\it Subaru} $z$, KPNO 4-m $B_WRIJHK_S$ and (deeper) {\it Spitzer} IRAC photometry. Our AGN sample is selected from the full photometric sample following the three criteria outlined in \citet{duncan2018a}, which expands on the previous work in this field by \citet{hickox2009}. 
Optical AGNs are selected based on their spectroscopic classification and/or their optical morphology and colours.
Infrared selected AGNs are selected using the {\it Spitzer} Deep Wide-Field Survey \citep[SDWFS;][]{ash09} and the \citet{donley2012} mid-infrared selection criteria.
Finally, X-ray AGNs are selected from the XBo\"otes survey \citep{murray2005}, a wide-field {\it Chandra} survey with (typical) integration times of $5~{\rm ks}$ per pointing and an soft X-ray (0.5 to 2 keV) flux limit of $\approx 5\times10^{-15}$~erg~ cm$^{-2}$~s$^{-1}$. 

\begin{figure*}
\centering
 \includegraphics[width=\textwidth]{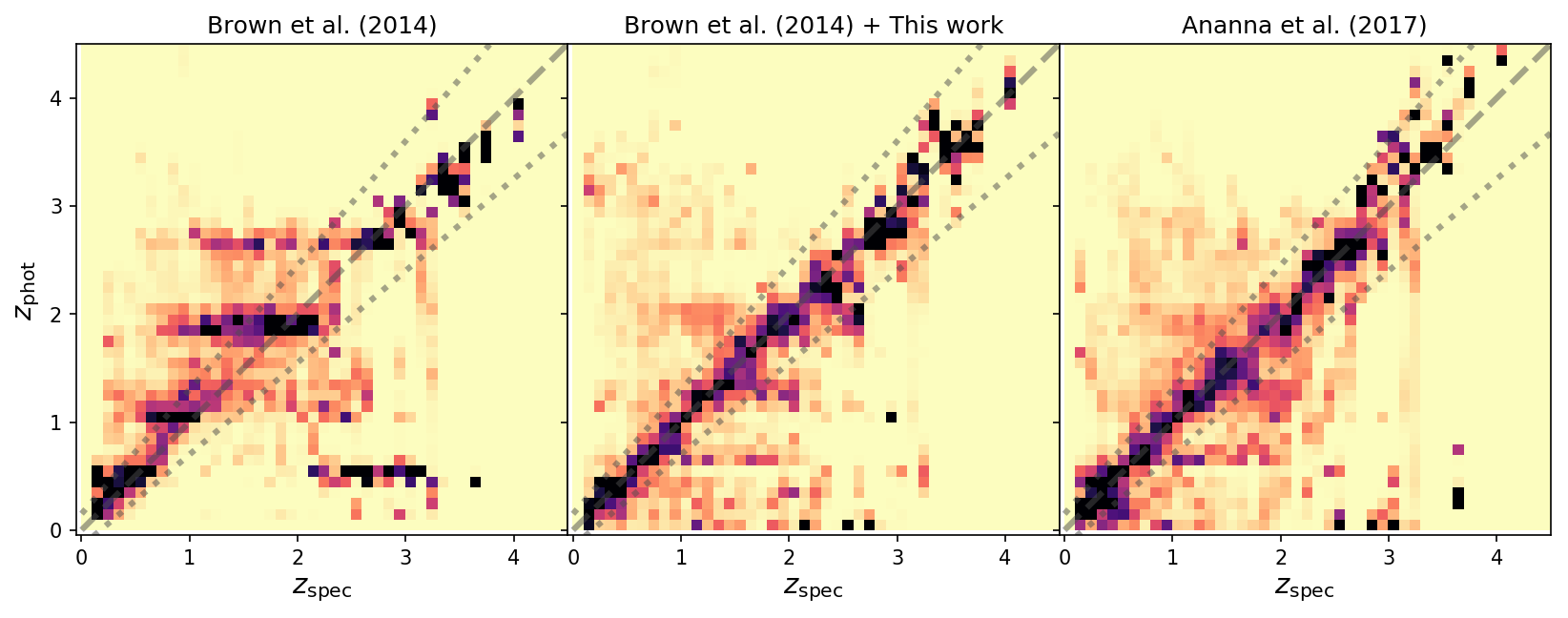}
  \caption{Spectroscopic vs photometric redshift comparison for the Bo\"{o}tes X-ray selected AGN sample at $0 < z < 4.5$ using three different template sets. Unsurprisingly, \citet{bro14} has the worst performance for AGN photometric redshifts as it (by construction) excludes powerful AGNs. Relative to photometric redshifts determined using the \citet{ananna2017} templates, photometric redshifts using our new templates have lower $\chi^2$ values and slightly better performance.}
 \label{fig:zspec_zphot}
\end{figure*}

Limiting the AGN sample to sources with spectroscopic redshifts, our final sample consists of 2058 sources in the redshift range $0 < z \leq 6.12$ with a median redshift of $z = 1.25$. The bulk of our spectroscopic redshifts come from the MMT AGN and Galaxy Evolution survey \citep[AGES; ][]{koc12}, although some additional redshifts come from {\it Keck}, {\it Gemini} and other observing programs. As the Bo\"{o}tes field has a $9~{\rm deg}^2$ area and AGES has an AB magnitude limit of $I=21.9$ for {\it Spitzer} IRAC selected AGNs, we have more luminous quasars and Seyferts than deeper surveys of smaller fields.

To compare the utility of our SEDs with prior literature, we have produced three sets of photometric redshifts. We have produced photometric redshifts using 1) the \citet{bro14} SEDs, which do not include luminous AGNs, 2) the AGN and galaxy SED libraries of \citet{ananna2017} and 3) our new AGN SEDs along with the SEDs of \citet{bro14}. 
Photometric redshifts for each set of templates were determined using \textsc{EAzY} \citep{brammer2008}. For simplicity, in all cases we apply the same template-fitting approach: using single-template fits only, including the calculation of photometric zero-point offsets, and incorporating an additional 10\% flux uncertainty (added in quadrature). We do not apply any magnitude priors when calculating the redshift posteriors at this stage.

When fitting in \textsc{EAzY}, we use the \citet{ananna2017} templates in a way that best matches their implementation with \textsc{LePhare} \citep{arnouts1999} for Stripe 82. Specifically, we apply dust extinction in the range of $0 \leq E(B-V) \leq 0.5$ following an SMC-like extinction curve \citep[specifically the parametrisation of][for this analysis]{pei1992}. For the \citet{bro14} templates, when fitted by themselves or when combined with the templates in this work, we apply extinction in the same fashion as for \citet{ananna2017}. However, for the AGN templates added in this work, we use a reduce range of extinction $0 \leq E(B-V) \leq 0.2$. For consistency, all templates are interpolated onto the same  wavelength grid when dust attenuation is applied\footnote{We caution that in addition to the SEDs having a varying step size in $\lambda$, the long length of some of the SEDs (e.g., $>10,000$ data points) may clash with the default setup of some photometric redshift codes.}.  

Formally, the predicted photo-$z$ posterior should account for both the likelihood of a source being a specific galaxy type (or template) at a redshift given the photometry \emph{and} the prior likelihood of that galaxy type existing at that redshift.
Improved photo-$z$ can therefore be gained by accounting for the relative number-density of different galaxy types as a function of redshift; particularly in the case of quasars and AGN where specific types can be extremely rare (e.g. BL Lac objects)\footnote{That said, only 12 of the 2058 AGNs in Bo\"{o}tes with spectroscopic redshifts are best-fit by the BL Lac templates and these objects are photometric redshift outliers regardless of whether the BL Lac templates are included in the library or not.}.
This improvement is illustrated by the photo-$z$ quality for X-ray selected AGN gained when removing AGN types typically too rare to be found in deep pencil-beam surveys \citep[e.g.,][]{salvato2009,hsu2014,salvato2018}.
However, we caution that for these tests we have used all of the templates provided and have not attempted to optimise performance using a subset of the templates, therefore minimising any potential systematics that may be induced by the photo-$z$ methodology.
We note that the depth of the X-ray survey that provides our primary test sample is well matched to that used in \citet{ananna2017}.
We are therefore testing a similar parameter space to that for which the \citeauthor{ananna2017} library has been optimised.

In Figure~\ref{fig:zspec_zphot} we present the photometric redshifts of Bo\"{o}tes field X-ray AGNs as a function of spectroscopic redshift.
In bins of spectroscopic redshift, we plot the stacked photo-$z$ posteriors (with no additional uncertainty calibration applied) - with the distributions normalised in bins of constant photometric redshift.
To improve the visual clarity due to the sparse sample of spectroscopic redshifts at $z_{\textup{spec}} > 2$, we stack the posteriors in broader redshift bins than used during template fitting.

Unsurprisingly, as \cite{bro14} does not include powerful AGNs, it performs poorly and has a significant number of catastrophic outliers. 
All three sets of photometric redshifts show some aliasing, with some erroneous photometric redshifts associated with strong emission lines moving in and out of particular filters. 
At first glance the photometric redshifts determined with our new templates and the templates of \citet{ananna2017} are of comparable quality. 
However, there are some subtler differences such as the templates from this work producing posteriors that are visibly more concentrated around the 1:1 trend at $z>2$.
Both the dramatic improvements over the previous \cite{bro14} estimates and the smaller improvements over the \citet{ananna2017} library are clearer when we investigate the statistical performance as a function of redshift. 

\begin{figure}
\centering
 \includegraphics[width=\columnwidth]{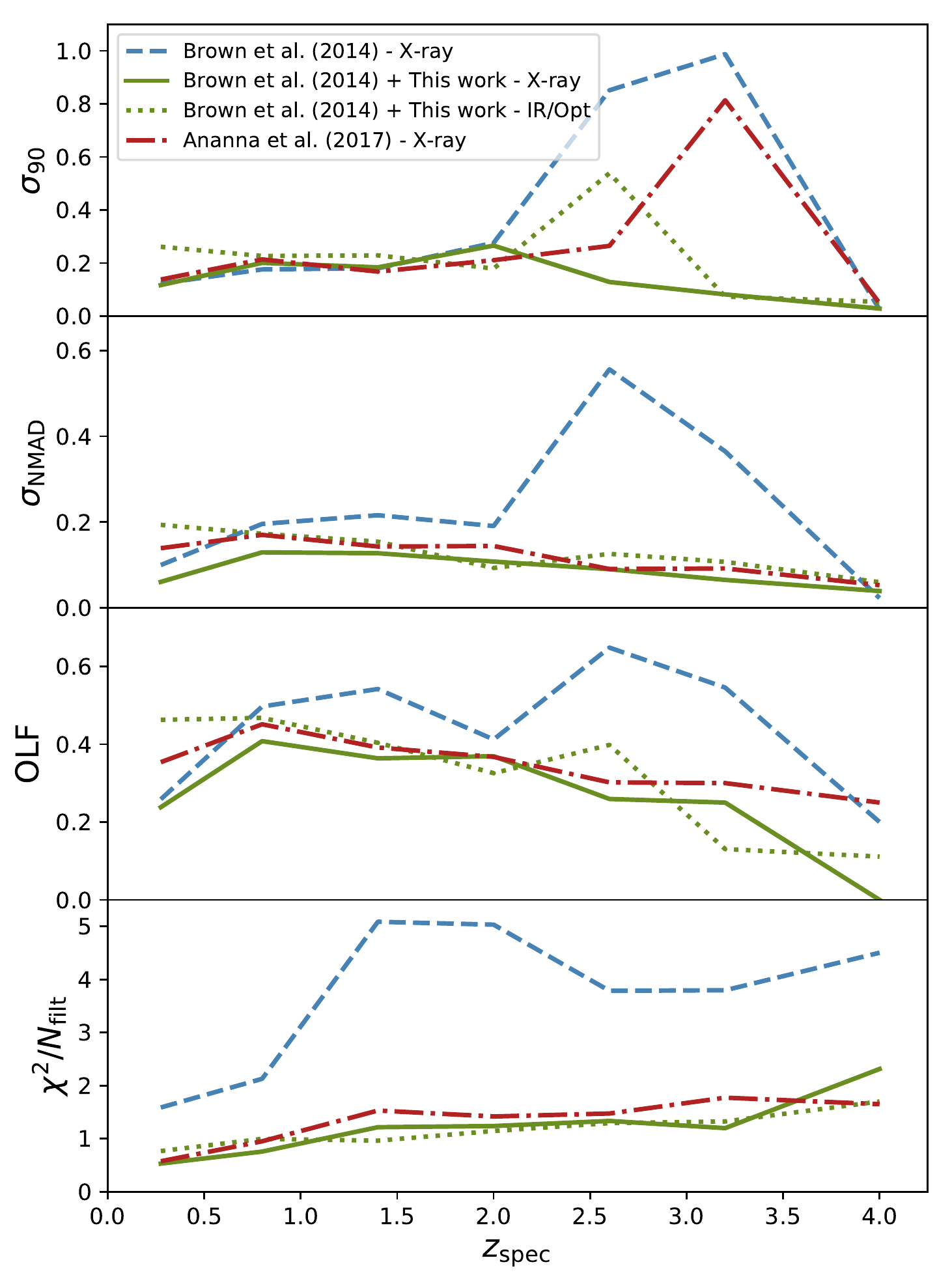}
  \caption{Photometric redshift quality metrics as a function of spectroscopic redshift for the three template sets (see text for metric definitions). Our new AGN SED templates, in combination with the \citet{bro14} galaxy SEDs, produce photometric redshifts for X-ray selected AGNs that are comparable to or better than the photometric redshifts produced using the \citet{ananna2017} SEDs.}
 \label{fig:zphot_stats}
\end{figure}

In Figure~\ref{fig:zphot_stats} we plot four metrics of photometric redshift quality: the 90\% clipped root mean squared scatter ($\sigma_{90}$), the normalised median absolute deviation ($\sigma_{\textup{NMAD}}$), the outlier fraction (OLF) and median $\chi^2$ divided by the number of filters. 
Our measure of robust scatter, $\sigma_{\textup{NMAD}}$, is defined as $1.48 \times \text{median} ( \left | \delta z \right | / (1+z_{\text{spec}}))$ as is commonly used within the photo-$z$ literature, where $\delta z = z_{\textup{spec}} - z_{\textup{phot}}$.
Similarly, we define the outlier criterion as $\left | \delta z \right | / (1+z_{\textup{spec}}) > 0.15$. 
When calculating these metrics, $z_{\textup{phot}}$ is chosen to be the maximum a-posteriori value of the redshift posterior (corresponding to the minimum $\chi^2$ given our assumed flat prior).
Finally, the bottom panel of Figure~\ref{fig:zphot_stats} shows the median  $\chi^{2}/N_{\textup{filt}}$ for sources in each redshift bin to illustrate the overall goodness-of-fit for each template set.
For the metrics presented in the top three panels of Figure~\ref{fig:zphot_stats}, sources with the worst 5\% of $\chi^{2}/N_{\textup{filt}}$ were excluded for each template set.

For all metrics presented in Figure~\ref{fig:zphot_stats}, we find our SEDs produce photometric redshifts for $z<2.5$ X-ray AGNs that are comparable to those using the \citet{ananna2017} SEDs, with improvements for $z>2.5$ AGNs including lower $\sigma_{90}$ and OLF.
We also find that our new SEDs are able to perform equally well for mid-infrared and optically selected AGN population - illustrating the wide range of parameter space represented in our SED library.
The templates presented in this work are able to produce consistently lower $\chi^{2}$ than either comparison template library, with a median $\chi^{2}/N_{\textup{filt}} \approx 1$ at all $z_{\textup{spec}}$; illustrating that we are able to better represent the observed optical to mid-IR colour space of high redshift AGNs than previous libraries.

\begin{figure}
\centering
 \includegraphics[width=\columnwidth]{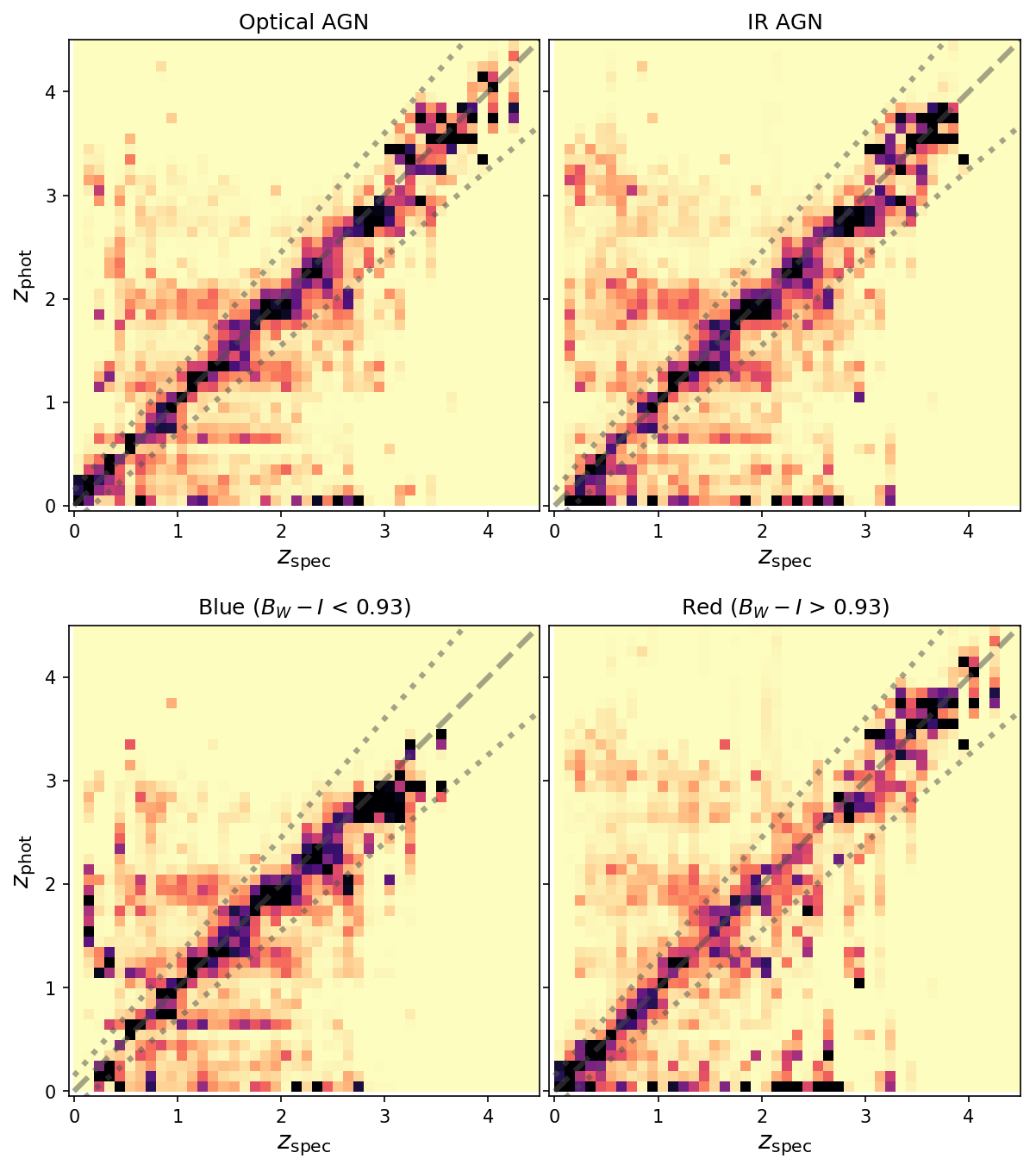}
  \caption{Photometric redshifts, determined with our SEDs, for AGNs that meet optical selection criteria, {\it Spitzer} infrared selection criteria, and when separated into blue and red apparent optical colours. As in Figure~\ref{fig:zspec_zphot} plot the stacked photo-$z$ posteriors, with the distributions normalised in bins of constant photometric redshift. While photometric redshift performance will inevitably depend on AGN selection and colour (and this can be seen above), the photometric redshift performance is comparable for the four subsamples shown.}
 \label{fig:agnsubsets}
\end{figure}

As AGN SEDs are a function of both selection criteria and (by definition) colour, we expect photometric redshift performance to also depend on selection criteria and colour. 
In Figure~\ref{fig:agnsubsets} we present the photometric redshifts, determined with our SEDs, as a function of spectroscopic redshift for AGNs that meet optical selection criteria, {\it Spitzer} infrared selection criteria \citep{donley2012}, an optical blue colour criterion ($B_W-I<0.93$) and an optical red colour criterion ($B_W-I>0.93$). 
Overall photometric redshift performance for these different subsets of AGNs is comparable, although red AGN photometric redshifts outperform blue AGN photometric redshifts over certain redshift ranges (and vice versa). We suspect the variation of photometric redshift performance is due to the contribution or absence of strong spectral features such as the 4000~\AA\, break in AGN host galaxies and the $\sim 1~{\rm \mu m}$ inflection in blue quasars SEDs. 

\begin{figure}
\centering
 \includegraphics[width=\columnwidth]{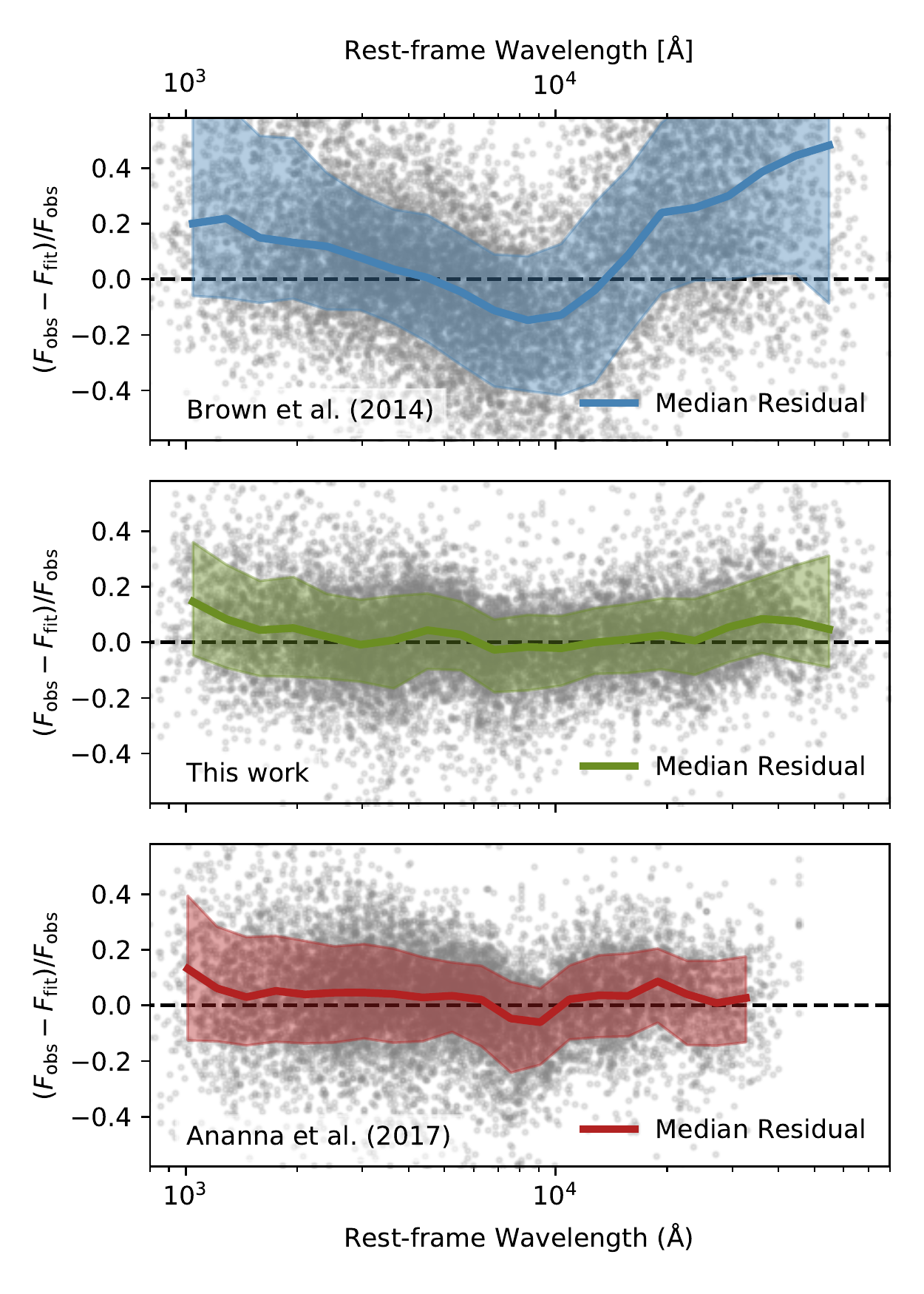}
  \caption{Rest-frame residuals for the template fits (with redshifts are fixed to the known spectroscopic redshift) using each of the three SED libraries employed in this work. Background grey points correspond to individual datapoints (i.e. one for each fitted filter per source) while the solid coloured lines correspond to the median residual within a given wavelength bin and the shaded region corresponds to the 16 and 84th percentiles (i.e., $\pm 1\sigma$). Our AGN SEDs in combination with the \citet{bro14} galaxy SEDs (middle panel) span a broad wavelength range while producing small residuals, with just 3\% of the residuals being $3\sigma$ outliers.}
 \label{fig:restframe_residuals}
\end{figure}

To further illustrate how our new library is able to better represent the panchromatic colours of the observed AGN population, in Figure~\ref{fig:restframe_residuals} we show the rest-frame residual fitting errors when the redshifts are fixed to the known spectroscopic redshift for our \emph{full} AGN sample.
Compared to the \citet{bro14} library alone, biases in the model colours are reduced by a factor of $\sim 2$ to 5 depending on the rest-frame wavelength.
The scatter in residual colours is also significantly reduced for the SEDs provided in this work compared to both the \citet{bro14} and \citet{ananna2017}; indicative of reduced model uncertainties.
Over rest-frame wavelengths of $0.1$ to $3~{\rm \mu m}$, the median absolute residual reduces from 18\% for the \citet{bro14} library to 9\% for the combined empirical library of this work. The \citet{ananna2017} SEDs have median absolute residual of 11\%, which is only slightly higher than our work although the \citet{ananna2017} SEDs are fitted to a more limited wavelength range.

\begin{figure}
\centering
 \includegraphics[width=\columnwidth]{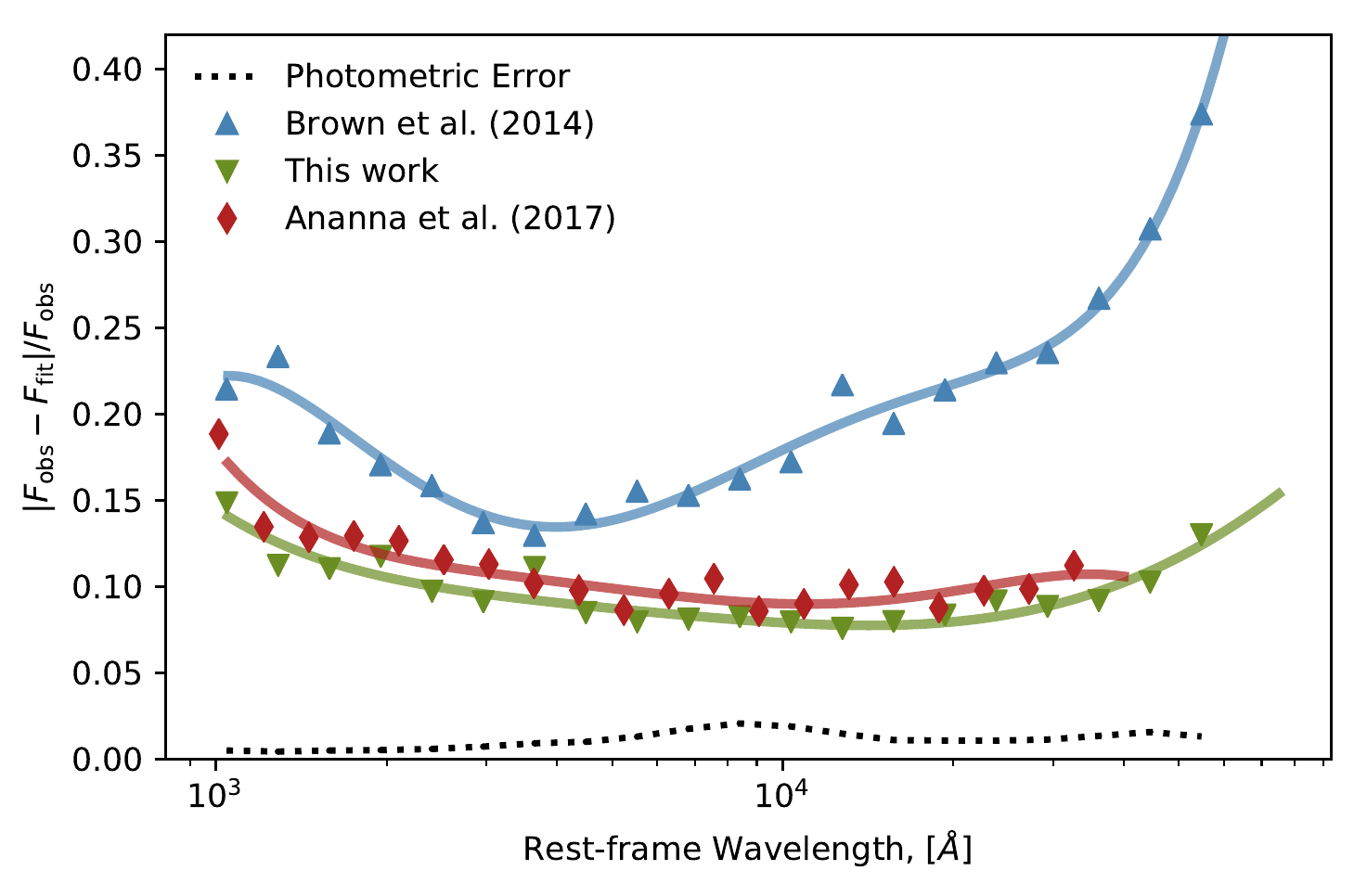}
  \caption{The template error function for each set of template SEDs (solid coloured lines), calculated by subtracting in quadrature the scaled average fractional error (black dotted line) from the median absolute rest-frame residuals (coloured symbols).}
 \label{fig:template_errors}
\end{figure}

In Figure~\ref{fig:template_errors} we highlight this reduction in model uncertainties by plotting the `template error function' as presented in \citet{brammer2008} for each set of SEDs.
For each set of SEDs we calculate the template error function by subtracting the scaled average fractional error from the median absolute rest-frame residual (in quadrature).
We find that for all three sets of template SEDs, the template error is highest for rest-frame UV and mid-IR emission - consistent with the pattern observed in \citet{brammer2008} for normal galaxies.
In \citet{brammer2008}, the increased uncertainty in these regimes is attributed to intrinsic variation in dust attenuation curves in restframe ultraviolet and the lack of mid-infrared emission from dust in their templates (our SEDs include include mid-infrared emission from dust and, where applicable, PAHs). 

The template error estimated from the fits incorporating the AGN SEDs in this work is largely comparable to that of the \citet{ananna2017} library. 
As seen in Figure~\ref{fig:template_errors}, the improvement in overall residuals discussed above can be attributed to improvement in the templates at rest-frame wavelengths of $<2000 \textup{\AA}$ and at $\sim1~{\rm \mu m}$.
Additionally, the substantial reduction in template error at $\lambda > 1~{\rm \mu m}$ relative to that of the \citet{bro14} library alone ($2\times$ lower) further illustrates how our new SEDs are able to better represent the observed range of mid-IR colours of AGNs. 

As illustrated by \citet{brammer2008}, the inclusion of the template error function when performing template fits (added in quadrature to the flux measurement errors) can yield improvements in the photo-$z$ estimates.
To verify that incorporating more realistic model uncertainties in the template fitting leads to further improvement in the photo-$z$s, we calculate a new set of estimates incorporating the template error function specific to each SED library (as presented in Figure~\ref{fig:template_errors}).
Additionally, we also incorporate a simple magnitude dependent prior using the observed $3.6\mu\textup{m}$ \emph{Spitzer} IRAC magnitude \citep[following the method outlined in][]{duncan2018a}.
For the semi-`optimised' photo-$z$ estimates using the extended SED library of this work, we find that the robust scatter reduces from $\sigma_{\textup{NMAD}} = 0.12$ to $\sigma_{\textup{NMAD}} = 0.095$ across all AGN types, while the overall outlier fraction drops substantially from 37\% to 31\%.

An exhaustive optimisation of the template fitting method for our AGN test sample is beyond the scope of this work and will be explored further in future studies.
However, the improvements gained from the minor changes outlined above illustrate the potential for further dramatic improvements in both precision and reliability.
For example, as demonstrated in the works of \citet{pol07}, \citet{salvato2009}, \citet{hsu2014} and \citet{ananna2017}, it is useful to account for the optical morphology of the source when choosing template libraries (e.g. excluding pure quasar SEDs for extended morphologies).
Similarly, applying an absolute magnitude prior can prevent severe degeneracies between different SEDs that would otherwise result in unphysical luminosities.  
Finally, given a particular scientific goal or target sample of interest, the exact subset of templates included in the fit or the range of dust attenuation allowed could be tuned based on our prior knowledge.

\section{Conclusions}
\label{sec:conclusions}

We have determined the $0.09$ to $30~{\rm \mu m}$ SEDs of 41 individual AGNs using a combination of archival spectroscopy, matched aperture photometry and models. In some instances the SEDs have been expanded into the X-ray, far-infrared and radio using combinations of archival spectroscopy, greybody models and simple empirical models. We have also produced 72 additional SEDs that  mimic Seyferts, by combining central SEDs of nearby Seyferts with galaxy SEDs from \citet{bro14}. All of the SEDs are available via DOI 10.17909/t9-3dbt-8734. 

To produce continuous SEDs using archival data taken over years (or decades), we multiplicatively scaled the individual spectra to produce continuous SEDs. For half the AGNs the scalings were between 0.5 and 2.0, but for 7 AGNs the scalings were below 0.33 or above 3. Some gaps in spectral coverage (or between spectra and models) were filled using simple models to interpolate between or extrapolate from the individual spectra. Comparison of the resulting SEDs with matched aperture photometry shows reasonable agreement, with most SEDs agreeing with photometry within a factor of 2. Offsets between the SEDs and photometry decrease with increasing wavelength (as expected given AGN variability) and some offsets are the result of host galaxy light impacting (large) matched aperture photometry.

The overall shapes of our SEDs are similar to those from the prior literature \citep[e.g.,][]{elvis1994} but we achieve broader wavelength coverage and higher spectral resolution, particularly in the X-ray, mid-infrared and far-infrared. For example, silicate emission features and far-infrared emission from warm dust are evident in our quasar templates. Optical and mid-infrared colours synthesised from our SEDs can differ significantly from those generated by galaxy and AGN SED libraries from the prior literature.

We tested the utility and precision of our AGN SEDs by using them to determine photometric redshifts for X-ray, optical and mid-infrared selected AGNs in the Bo\"{o}tes field. AGN photometric redshifts utilising  SED libraries without AGNs can have gross redshift errors. AGN photometric redshifts determined with our AGN SEDs and the \citet{bro14} galaxy SEDs have a typical  scatter of $\sigma_{\textup{NMAD}}=0.096 \times (1+z)$, median reduced $\chi^2$ values of $\simeq 1$, and typical flux density residuals of 9\% (with the exact value being a function of wavelength). We thus conclude that our empirical SEDs are functional models for many of the AGNs that are being (or soon will be) detected by the current generation of wide-field surveys, including WISE, LOFAR, ASKAP, eROSITA and SPHEREx. 

\section*{Acknowledgements}

A number of astronomers generously shared their knowledge and reduced spectra while we prepared this paper, including A. Barth, R. Decarli, M. Elvis, E. Glikman, R. Hickox, N. Hurley-Walker, M. Koss, J. Kuraszkiewicz, I. Lamperti, J. McDowell, C. Mundell, K. Oh, R. Riffel, A. Rodr{\'{\i}}guez-Ardila, D. Rupke, P. Smith, G. Snyder, P. F. Spinelli, D. Stern, C. Tadhunter, S. Veilleux, S. White, and B. Wilkes. M. Kirk undertook preliminary work for this paper as part of a Monash University PHS~3350 undergraduate research project. We thank M. Durr\'{e} for reducing the ESO SINFONI spectra used in this paper. 

Some of the data presented in this paper were obtained from the Mikulski Archive for Space Telescopes (MAST). STScI is operated by the Association of Universities for Research in Astronomy, Inc., under NASA contract NAS5-26555. This work is based in part on observations made with ESO Telescopes at the La Silla Paranal Observatory under programme IDs 60.A-9339(A), 091.B-0256(A), 091.B-0900(B), 095.B-0015(A), 097.B-0080(A), and 097.B-0640(A).


This research has made use of data and/or software provided by the High Energy Astrophysics Science Archive Research Center (HEASARC), which is a service of the Astrophysics Science Division at NASA/GSFC and the High Energy Astrophysics Division of the Smithsonian Astrophysical Observatory.

This work is based in part on observations made with the Galaxy Evolution Explorer (GALEX). GALEX is a NASA Small Explorer, whose mission was developed in cooperation with the Centre National d'\'{E}tudes Spatiales (CNES) of France and the Korean Ministry of Science and Technology. GALEX was operated for NASA by the California Institute of Technology under NASA contract NAS5-98034. The {\it Swift} UVOT was designed and built in collaboration between MSSL, PSU, SwRI, Swales Aerospace and GSFC, and was launched by NASA.

Funding for the Sloan Digital Sky Survey IV has been provided by the Alfred P. Sloan Foundation, the U.S. Department of Energy Office of Science, and the Participating Institutions. SDSS-IV acknowledges support and resources from the Center for High-Performance Computing at the University of Utah. SDSS-IV is managed by the Astrophysical Research Consortium for the 
Participating Institutions of the SDSS Collaboration.

The Pan-STARRS1 Surveys (PS1) have been made possible through contributions of the Institute for Astronomy, the University of Hawaii, the Pan-STARRS Project Office, the Max-Planck Society and its participating institutes, the Max Planck Institute for Astronomy, Heidelberg and the Max Planck Institute for Extraterrestrial Physics, Garching, The Johns Hopkins University, Durham University, the University of Edinburgh, Queen's University Belfast, the Harvard-Smithsonian Center for Astrophysics, the Las Cumbres Observatory Global Telescope Network Incorporated, the National Central University of Taiwan, the Space Telescope Science Institute, the National Aeronautics and Space Administration under Grant No. NNX08AR22G issued through the Planetary Science Division of the NASA Science Mission Directorate, the National Science Foundation under Grant No. AST-1238877, the University of Maryland, and Eotvos Lorand University (ELTE).

The national facility capability for SkyMapper has been funded through ARC LIEF grant LE130100104 from the Australian Research Council, awarded to the University of Sydney, the Australian National University, Swinburne University of Technology, the University of Queensland, the University of Western Australia, the University of Melbourne, Curtin University of Technology, Monash University and the Australian Astronomical Observatory. SkyMapper is owned and operated by The Australian National University's Research School of Astronomy and Astrophysics. The survey data were processed and provided by the SkyMapper Team at ANU. The SkyMapper node of the All-Sky Virtual Observatory (ASVO) is hosted at the National Computational Infrastructure (NCI). Development and support the SkyMapper node of the ASVO has been funded in part by Astronomy Australia Limited (AAL) and the Australian Government through the Commonwealth's Education Investment Fund (EIF) and National Collaborative Research Infrastructure Strategy (NCRIS), particularly the National eResearch Collaboration Tools and Resources (NeCTAR) and the Australian National Data Service Projects (ANDS).

This publication makes use of data products from the Two Micron All Sky Survey, which is a joint project of the University of Massachusetts and the Infrared Processing and Analysis Center/California Institute of Technology, funded by the National Aeronautics and Space Administration and the National Science Foundation.

This research is based in part on observations with {\it Akari}, a JAXA project with the participation of ESA. This work is based in part on observations made with the {\it Spitzer} Space Telescope, which is operated by the Jet Propulsion Laboratory, California Institute of Technology under a contract with NASA. This publication makes use of data products from the Wide-field Infrared Survey Explorer, which is a joint project of the University of California, Los Angeles, and the Jet Propulsion Laboratory/California Institute of Technology, funded by the National Aeronautics and Space Administration. 

This publication makes use of data products from the Wide-field Infrared Survey Explorer, which is a joint project of the University of California, Los Angeles, and the Jet Propulsion Laboratory/California Institute of Technology, funded by the National Aeronautics and Space Administration. This work is based in part on observations made with the {\it Spitzer} Space Telescope, which is operated by the Jet Propulsion Laboratory, California Institute of Technology under a contract with NASA.

Herschel is an ESA space observatory with science instruments provided by European-led Principal Investigator consortia and with important participation from NASA. Based in part on observations with ISO, an ESA project with instruments funded by ESA Member States (especially the PI countries: France, Germany, the Netherlands and the United Kingdom) and with the participation of ISAS and NASA.

The WMAP mission is made possible by the support of the Office of Space Sciences at NASA Headquarters and by the hard and capable work of scores of scientists, engineers, technicians, machinists, data analysts, budget analysts, managers, administrative staff, and reviewers

This scientific work makes use of the Murchison Radio-astronomy Observatory, operated by CSIRO. We acknowledge the Wajarri Yamatji people as the traditional owners of the Observatory site. Support for the operation of the MWA is provided by the Australian Government (NCRIS), under a contract to Curtin University administered by Astronomy Australia Limited. We acknowledge the Pawsey Supercomputing Centre which is supported by the Western Australian and Australian Governments.

This research has made use of NASA's Astrophysics Data System Bibliographic Services. This research has made use of the NASA/IPAC Extragalactic Database (NED) which is operated by the Jet Propulsion Laboratory, California Institute of Technology, under contract with the National Aeronautics and Space Administration.




\bibliographystyle{mnras}
\bibliography{agnsed2018} 



\clearpage 

\appendix

\section{Photometry}

\begin{table*}
	\centering
	\caption{Measured photometry (in mJy) corrected for Milky Way foreground extinction (the complete table can be found online).}
	\label{table:obsfluxes}
    \scriptsize
	\begin{tabular}{lccccccc}
\hline
Name &  $FUV_{GALEX}$  &  $UVW2_{\it Swift}$  &  $UVM2_{\it Swift}$  &  $NUV_{GALEX}$  &  $UVW1_{\it Swift}$  &  $U_{\it Swift}$  &  $    u_{SM}$ \\ 
     &            $  u_{SDSS}$  &  $    v_{SM}$  &  $B_{\it Swift}$  &  $  g_{SDSS}$  &  $   g_{PS1}$  &  $    g_{SM}$  &  $V_{\it Swift}$ \\ 
     &            $  r_{SDSS}$  &  $   r_{PS1}$  &  $    r_{SM}$  &  $  i_{SDSS}$  &  $   i_{PS1}$  &  $    i_{SM}$  &  $   z_{PS1}$ \\ 
     &            $  z_{SDSS}$  &  $    z_{SM}$  &  $   y_{PS1}$  &  $ J_{2MASS}$  &  $ H_{2MASS}$  &  $ K_{2MASS}$  &  $ W1_{WISE}$ \\ 
     &            $3.6_{IRAC}$  &  $4.5_{IRAC}$  &  $ W2_{WISE}$  &  $5.0_{IRAC}$  &  $8.0_{IRAC}$  &  $ W3_{WISE}$  &  $ W4_{WISE}$ \\ 
     &            $ 24_{MIPS}$  &  $ 70_{PACS}$  &  $100_{PACS}$  &  $160_{PACS}$  &  $250_{SPIRE}$  &  $350_{SPIRE}$  &  $500_{SPIRE}$ \\ 
\hline     
             3C 120  &     -  &  16.1  &  17.9  &     -  &  14.5  &  16.3  &     - \\ 
                     &     -  &     -  &  12.8  &     -  &  14.6  &     -  &  14.7 \\ 
                     &     -  &  22.0  &     -  &     -  &  18.9  &     -  &  16.9 \\ 
                     &     -  &     -  &  22.0  &  23.3  &  31.5  &  48.8  &  61.1 \\ 
                     &  53.8  &  63.4  &  84.4  &   131  &   141  &   219  &   561 \\ 
                     &   597  &  1294  &     -  &  1386  &   697  &   496  &   457 \\ 
             3C 273  &  24.1  &     -  &     -  &     -  &     -  &     -  &     - \\ 
                     &  36.4  &     -  &     -  &     -  &  27.3  &     -  &     - \\ 
                     &     -  &  25.3  &     -  &     -  &     -  &     -  &  27.6 \\ 
                     &  30.2  &     -  &  31.5  &  32.3  &  39.6  &  69.0  &   132 \\ 
                     &   175  &   190  &   178  &   241  &   277  &   277  &   514 \\ 
                     &   685  &     -  &     -  &     -  &  1081  &  1546  &  2201 \\ 
             3C 351  &  1.43  &     -  &     -  &  2.87  &     -  &     -  &     - \\ 
                     &  2.39  &     -  &     -  &  2.70  &  2.83  &     -  &     - \\ 
                     &  3.12  &  3.39  &     -  &  3.24  &  3.27  &     -  &  4.67 \\ 
                     &  4.42  &     -  &  3.84  &  3.70  &  4.43  &  7.25  &  14.9 \\ 
                     &  19.0  &  21.1  &  21.9  &  27.6  &  31.5  &  37.0  &   104 \\ 
                     &   114  &   190  &   165  &  84.1  &  56.3  &     -  &     - \\ 
\hline
\end{tabular}
\end{table*}

\begin{table*}
	\centering
	\caption{Synthetic photometry (in mJy) derived from the SEDs (the complete table can be found online).}
	\label{tab:synfluxes}
    \scriptsize
	\begin{tabular}{lcccccccc}
\hline
Name &  $FUV_{GALEX}$  &  $UVW2_{\it Swift}$  &  $UVM2_{\it Swift}$  &  $NUV_{GALEX}$  &  $UVW1_{\it Swift}$  &  $U_{\it Swift}$  &  $    u_{SM}$ \\ 
     &  $  u_{SDSS}$  &  $    v_{SM}$  &  $B_{\it Swift}$  &  $  g_{SDSS}$  &  $   g_{PS1}$  &  $    g_{SM}$  &  $V_{\it Swift}$ \\ 
     &  $  r_{SDSS}$  &  $   r_{PS1}$  &  $    r_{SM}$  &  $  i_{SDSS}$  &  $   i_{PS1}$  &  $    i_{SM}$  &  $   z_{PS1}$ \\ 
     &  $  z_{SDSS}$  &  $    z_{SM}$  &  $   y_{PS1}$  &  $ J_{2MASS}$  &  $ H_{2MASS}$  &  $ K_{2MASS}$  &  $ W1_{WISE}$ \\ 
     &  $3.6_{IRAC}$  &  $4.5_{IRAC}$  &  $ W2_{WISE}$  &  $5.0_{IRAC}$  &  $8.0_{IRAC}$  &  $ W3_{WISE}$  &  $ W4_{WISE}$ \\ 
     &  $ 24_{MIPS}$  &  $ 70_{PACS}$  &  $100_{PACS}$  &  $160_{PACS}$  &  $250_{SPIRE}$  &  $350_{SPIRE}$  &  $500_{SPIRE}$ \\ 
\hline     
             3C 120  &  15.9  &  13.6  &  12.7  &  12.8  &  13.1  &  13.4  &  13.6 \\ 
                     &  13.4  &  12.5  &  9.63  &  10.3  &  10.0  &  9.83  &  9.43 \\ 
                     &  10.6  &  13.2  &  13.0  &  10.7  &  10.1  &  9.94  &  10.7 \\ 
                     &  10.7  &  10.7  &  10.5  &  18.5  &  29.3  &  50.3  &  88.3 \\ 
                     &  93.6  &   109  &   111  &   126  &   164  &   302  &   676 \\ 
                     &   695  &  1378  &  1487  &  1219  &   755  &   469  &   606 \\ 
             3C 273  &  26.1  &  30.6  &  31.2  &  31.4  &  32.6  &  34.5  &  33.6 \\ 
                     &  33.6  &  33.5  &  31.4  &  31.1  &  30.9  &  31.4  &  33.1 \\ 
                     &  30.6  &  30.7  &  30.5  &  38.2  &  38.7  &  39.9  &  30.1 \\ 
                     &  30.6  &  31.1  &  31.8  &  36.3  &  50.0  &  84.0  &   150 \\ 
                     &   158  &   188  &   190  &   214  &   257  &   344  &   641 \\ 
                     &   652  &   807  &   841  &   919  &  1167  &  1751  &  2841 \\ 
             3C 351  &  1.55  &  1.98  &  2.18  &  2.14  &  2.21  &  2.70  &  2.67 \\ 
                     &  2.75  &  2.97  &  2.96  &  3.00  &  3.03  &  3.04  &  3.13 \\ 
                     &  3.50  &  3.62  &  3.66  &  3.74  &  3.69  &  3.65  &  5.05 \\ 
                     &  4.82  &  5.13  &  3.96  &  4.15  &  5.44  &  8.69  &  19.1 \\ 
                     &  20.7  &  25.8  &  26.1  &  28.2  &  31.2  &  45.1  &   115 \\ 
                     &   122  &   184  &   154  &  93.8  &  49.0  &  28.2  &  38.4 \\ 
\hline
\end{tabular}
\end{table*}


\bsp	
\label{lastpage}
\end{document}